\pdfoutput=1
\documentclass[article,aps,nofootinbib,twocolumn,superscriptaddress]{revtex4-1}
\input epsf
\usepackage{graphics}
\usepackage{amsmath}
\usepackage{amssymb}
\usepackage{bm}
\usepackage{braket}
\usepackage{booktabs}
\usepackage{subfigure}
\usepackage{subfloat}
\usepackage{float}
\usepackage[usenames,svgnames]{xcolor}
\usepackage{hyperref}

\hypersetup{
    colorlinks=true,
    urlcolor=SteelBlue,
    linkcolor=red,
    citecolor=blue,
}

\usepackage{color}
\usepackage{dcolumn}
\usepackage{hyphenat}

\def\be{\begin{equation}}
\def\ee{\end{equation}}
\def\ba{\begin{eqnarray}}
\def\ea{\end{eqnarray}}

\usepackage{graphicx}

\begin{document}

\title{Observational Constraints on Constant Roll Inflation}

\author{Jos\'e T. G\'alvez Ghersi}
\email{joseg@sfu.ca}
\author{Alex Zucca}
\email{azucca@sfu.ca}
\author{Andrei V. Frolov}
\email{frolov@sfu.ca}\affiliation{Department of Physics, Simon Fraser University, Burnaby, BC, V5A 1S6, Canada}
\date{\today}

\begin{abstract}
Constant-roll inflation was recently introduced by Motohashi, Starobinsky and Yokoyama as a phenomenological way to parametrize deviations from the slow-roll  scenarios.
 In this paper, we investigate the dynamics of both the background and the perturbations in this model, without making any slow-roll assumptions. 
 The perturbation spectra are computed with an efficient and accurate novel method that allowed us to quickly scan the parameter space of constant-roll inflation. 
 We  derive the constraints on the model parameters from the cosmic microwave background anisotropy measurements provided by the joint analysis of the Planck Collaboration and the BICEP2/Keck Array data.
\end{abstract}

\maketitle

\section{Introduction}
\label{sec:Introduction}
The inflationary paradigm has been one of the most successful developments since its very first appearance in \citep{Starobinsky:1980te, Sato:1981, PhysRevD.23.347, Linde:1982, Albrecht:1982}, as its characteristic accelerated expansion solves most of the caveats of standard big bang cosmology.
In its most common form, inflation is driven by a scalar degree of freedom rolling slowly down a not very steep potential. Quantum fluctuations of the inflaton source the primordial inhomogeneities from which the actual large scale structure of the Universe emerges. 

Throughout the years, a multitude of inflationary models were proposed where the dynamics of the background field is highly overdamped, and the production of scalar and tensor fluctuations can be completely characterized by the so-called slow-roll parameters. Slow roll by itself is not a necessary condition for an inflationary model to be viable, and it is interesting to also explore models which break away from the slow-roll restrictions.
Motohashi \textit{et.\ al.}\ recently introduced a constant-roll inflation \citep{Motohashi:2014ppa,Motohashi:2017aob} which replaces the usual slow-roll condition with an ansatz that the field rolls at a constant rate, be it slow or not. The model is rather neat as it characterizes the deviation from a slow roll by a single parameter and allows analytic integration of the expansion history and the full scalar field potential reconstruction. The potential driving constant-roll inflation only differs from the one in natural inflation \citep{PhysRevLett.65.3233} by the addition of a negative cosmological constant (with specially chosen value). The constant-roll rate can be tuned by the period of the potential, which corresponds to the global symmetry breaking scale in natural inflation. In this paper, we derive constraints on constant-roll inflation from the cosmic microwave background (CMB) anisotropies data \citep{Adam:2015rua, Ade:2015xua, Ade:2015lrj, Ade:2015tva}.

In previous efforts \citep{Motohashi:2017aob}, the confrontation of this model with observational data used the well-known consistency relations between slow-roll and fluctuation spectra parameters \citep{Liddle:1994dx, Gong:2001he}, which assumes that using the exact background solutions, one can describe certain features of the perturbations spectra by using slow-roll approximation while this holds accurately in the parameter range to be explored. Even though this assumption is not inconsistent with the data, we find that there are small but noticeable differences if one computes the fluctuation spectra exactly. In our approach we do not impose any slow-roll assumptions and instead just directly evaluate the scalar and tensor power spectra of primordial fluctuations in a sufficiently large sector of the parameter space by numerical integration. This procedure allows us to evaluate deviations from the standard slow-roll approximation of the spectral index $n_s$ and the tensor-to-scalar ratio $r$. Achieving this with adequate parameter sampling and high precision throughout the mode evolution can be computationally expensive. We use a single-field version of our general method \citep{Ghersi:2016gee} which separates the fast and slow scales in the mode evolution to exponentially increase efficiency of sub-horizon integration. Our computational method allows us to scan a significant portion of the parameter space quickly and extremely accurately on a personal computer with minimal specifications.            

Constant-roll inflation has an uncertainty on the field value where inflation ends, as the potential needs to be cut at some value $\phi_0$ to get an exit from inflation. This introduces a third parameter to the model in addition to the mass scale and the roll rate, which fortunately turns out to be entirely degenerate as far as fluctuation spectra are concerned. This allows us to set tight constraints on two combinations of the model parameters for constant-roll inflation: one that determines the amplitude of scalar and tensor perturbations (along with the characteristic energy scale where inflation occurs), and the second one which sets the roll rate (and quantifies the deviations from the slow-roll approximation). We also compare constant-roll inflation with other models via estimation of the allowed region on the $r$ versus $n_s$ diagram, where each point can be identified with (at least) one choice of the model parameters after the spectrum is evaluated at the pivot scale. 

\begin{figure*}
\centering
\subfigure{
\includegraphics[width=.45\textwidth]{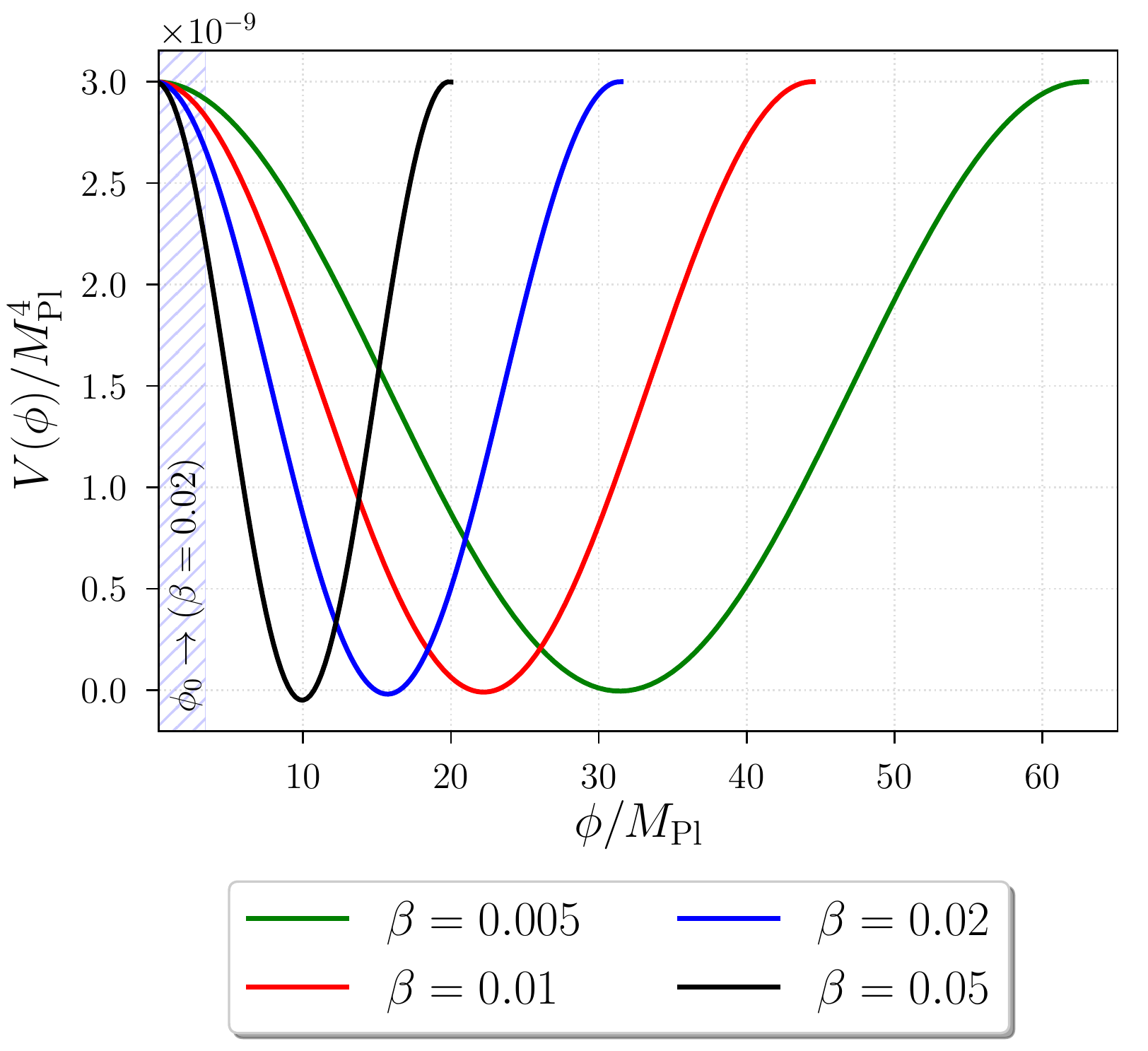}} \,
\subfigure{
\includegraphics[width=.5\textwidth]{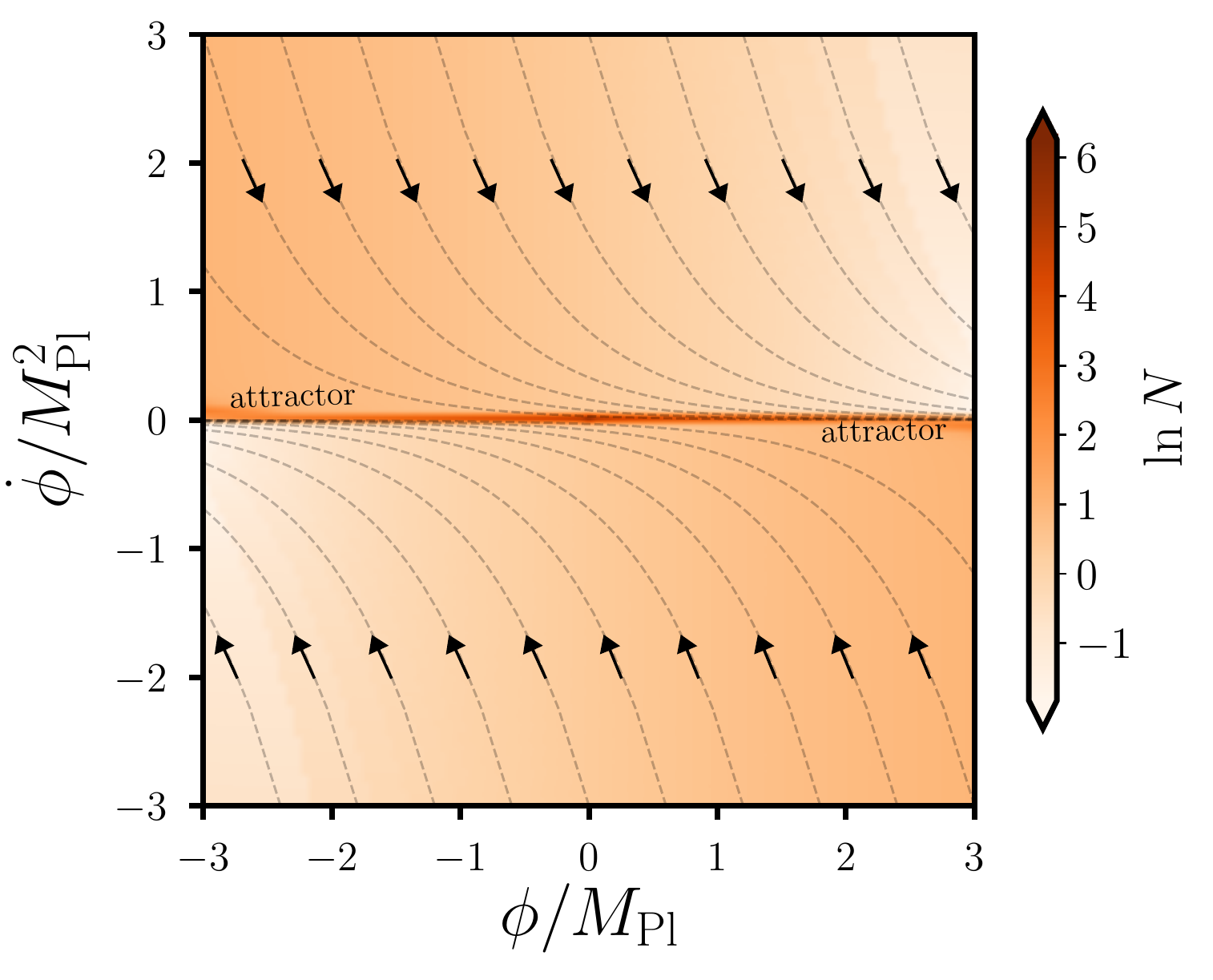}}
\caption{\label{fig:background} Left panel: constant-roll inflation potential \eqref{eq:CR_potential} for the model parameters $M^2=2.0\times10^{-9}M_{\mathrm{Pl}}^2$ and different values of $\beta$. The region of interest is in the range $\phi\in (0;\phi_0]$, shown for the specific value of $\beta=0.02$. Right panel: Map of initial conditions in the phase space for $M^2=2.0\times10^{-9}M_{\mathrm{Pl}}^2$, $N_*=55$ and $\beta=0.02$. The color map represents the number of e-folds before reaching $\pm\phi_0$. Phase space trajectories converge as a power law towards the attractor (instead of exponentially, as is usually the case in slow-roll inflation) as in the case of power law inflation, which is a particular scenario of the constant-roll model. } 
\end{figure*}

The layout of this paper is as follows: In section \ref{sec:Model_Background}, we present the model and scan a representative subset of the background phase space in order to determine the expansion history due to each choice of initial conditions. Our exploration of the phase space also includes the attractor formed by converging field trajectories. We describe the dynamics of the scalar and tensor fluctuations in section \ref{sec:Perturbations}, where we discuss the separation technique of scalar and tensor modes into fast and slow components, a mode injection scheme to calculate the spectra numerically, and compute the representative spectra given a set of arbitrary model parameters. We show that both the scalar and tensor spectra are featureless, and, as a consistency check, we also explicitly show that none of the modes evolve on super-horizon scales. In section \ref{sec:Observational_Constraints}, we use the joint likelihood data from Planck 2015 \citep{Adam:2015rua, Ade:2015xua, Ade:2015lrj} and BICEP2/Keck Array \cite{Ade:2015tva} to constrain the constant-roll inflation model parameters\footnote{We are profoundly aware of the Planck 2018 release \cite{Akrami:2018vks, Aghanim:2018eyx, Akrami:2018odb}, however the joint likelihood with BICEP/Keck is not available yet at the time of this writing.}. Finally, in section \ref{sec:Discussions}, we discuss the results and conclude.

\section{Model and Background Dynamics}
\label{sec:Model_Background}

\begin{figure*}[tbh]
\centering
\subfigure{
\includegraphics[width=.45\textwidth]{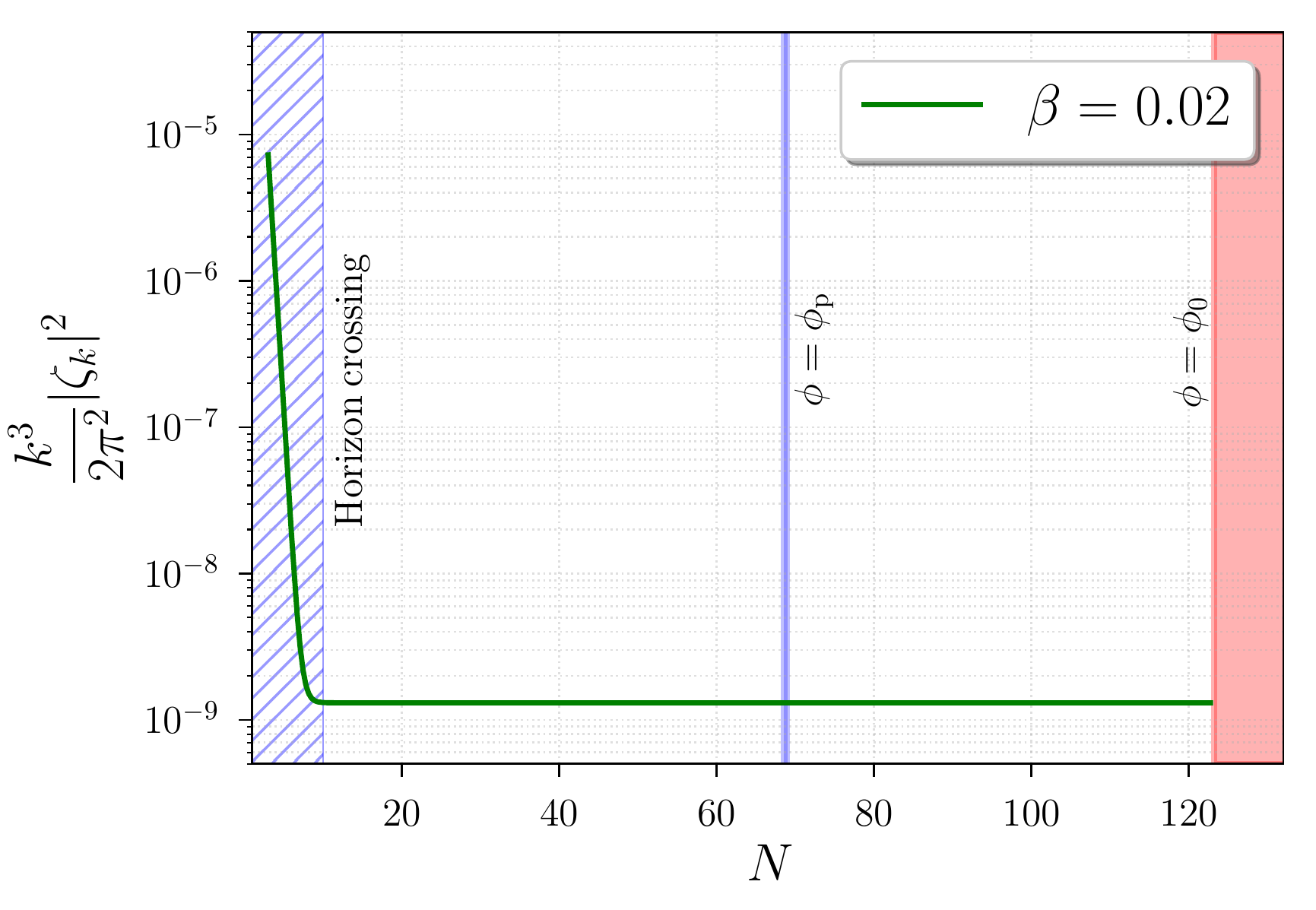}} \,
\subfigure{
\includegraphics[width=.465\textwidth]{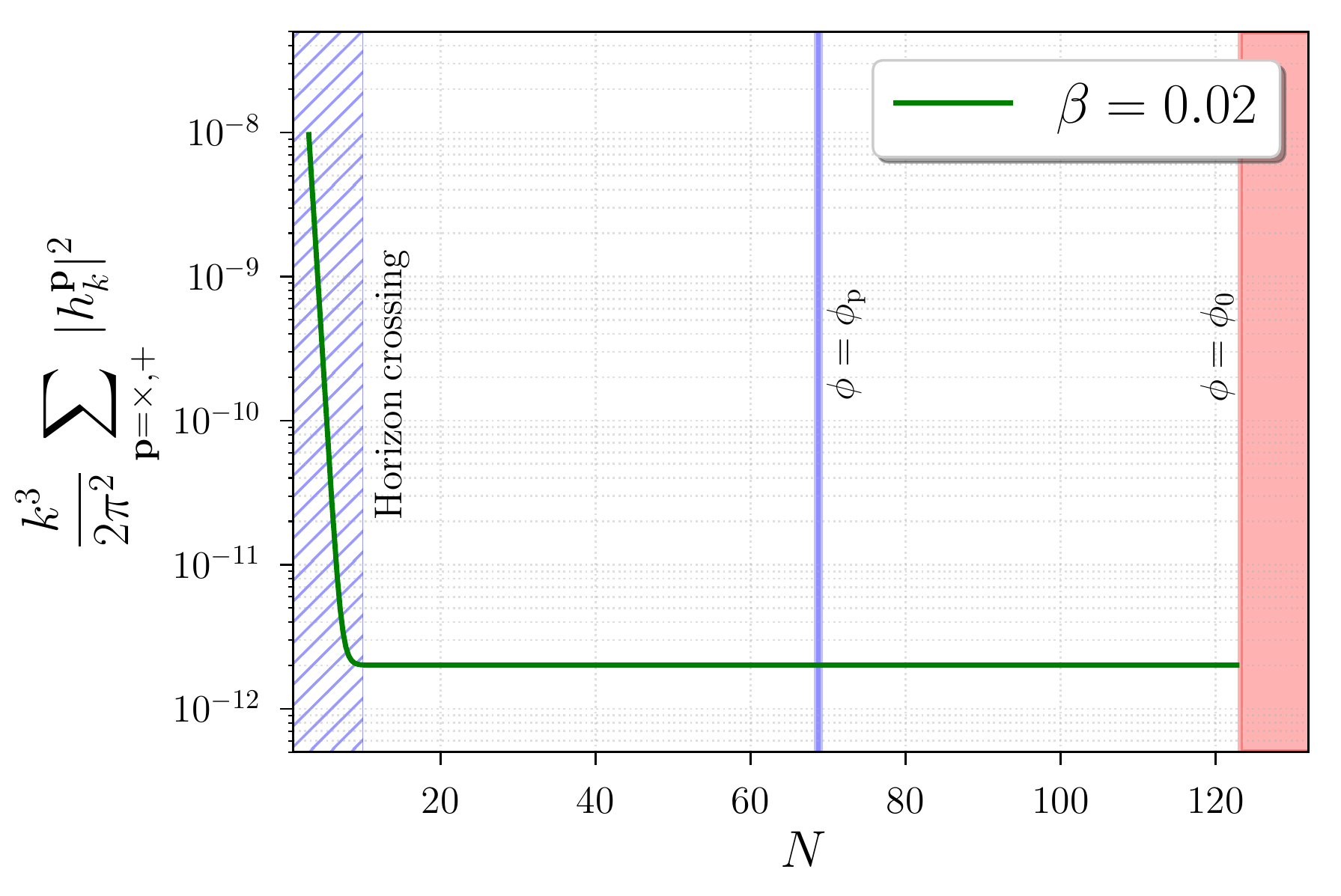}}
\caption{\label{fig:pert_evol} Left panel: Example of the evolution of the curvature fluctuations for $\beta=0.02$, $N_*=0$ and $M^2=10^{-11}M_{\mathrm{Pl}}^2$. Right panel: Evolution of the corresponding sum of the two tensor polarizations for the same model parameters. $N$ is the number of e-folds from the start of numerical evolution of the background.  Using the definition in \eqref{eq:phi_p}, both figures also include the instant in which the CMB pivot modes leave the horizon. We show extremely long wavelength modes emerging deep from the sub-horizon scales to demonstrate that there is absolutely no evolution on super-horizon scales in all cases.} 
\end{figure*}

In this section, we review the results of \citep{Motohashi:2017aob} using one of the forms of the potential reconstructed in \citep{Motohashi:2014ppa}. These are necessary to provide a full description of the background evolution of this model. The dynamics of the inflaton field minimally coupled to gravity is governed by the effective action
\begin{equation}
S=\displaystyle{\int d^4x\sqrt{-g}\left[\frac{M_{\mathrm{Pl}}^2}{2}R-\frac{1}{2}\,g^{\mu\nu}\partial_\mu\phi\,\partial_\nu\phi-V(\phi)\right]},
\label{eq:action}
\end{equation} 
where $g \equiv \det g_{\mu \nu}$,  $M_{\rm Pl} = (8 \pi G)^{-1/2}$ is the reduced Planck mass and $\phi$ is the inflaton field with a canonical kinetic term. We will use the signature $(-,+,+,+)$ and will assume the spatially flat Friedmann-Lema{\^\i}tre-Robertson-Walker metric for background. The constant-roll inflation potential $V(\phi)$ we use throughout this paper was derived in \citep{Martin:2012pe, Motohashi:2017aob} after reducing the order of the standard background equation of motion
\begin{equation}
\ddot{\phi}+3H\dot{\phi}+V'(\phi)=0
\label{eq:back_eq_mov}
\end{equation} 
by the constant-roll ansatz $\ddot{\phi}=\beta H\dot{\phi}$. The role of $\beta$ is to parametrize the magnitude of the second time derivative and thus, the deviations from the slow-roll approximation. Using the two Friedmann equations
\begin{eqnarray}
&3M_{\mathrm{Pl}}^2H^2=\frac{\dot{\phi}^2}{2}+V(\phi),\nonumber\\
&-2M_{\mathrm{Pl}}^2\dot{H}=\dot{\phi}^2,
\label{eq:Friedmann_eqns}
\end{eqnarray} 
it is possible to find a particular solution for the background evolution and reconstruct the constant-roll inflation potential $V(\phi)$, which turns out to be
\begin{equation}
V(\phi)=3M_{\mathrm{Pl}}^2M^2\left[1-\frac{3+\beta}{6}\bigg\{1-\cos\left(\frac{\sqrt{2\beta}\phi}{M_{\mathrm{Pl}}}\right)\bigg\}\right].
\label{eq:CR_potential}
\end{equation}
The shape of the potential is illustrated in the left panel of Fig.~\ref{fig:background} and evaluated at different values of $\beta$. The mass $M$ determines both the energy scale at which inflation occurs and the amplitude of the primordial fluctuations. The potential \eqref{eq:CR_potential} can become negative, and must be cut off somewhere before that to exit the inflation gracefully.

In the absence of an inherent point on the potential where inflation ends, it is important to explicitly specify the field range where we will evaluate the curvature and tensor fluctuations. The background field evolution proceeds from arbitrarily small values to some upper bound $\phi_0$ where potential is modified and inflation ends, which we parametrize by setting to be $N_*$ e-folds away from reaching the critical point where $V=0$ in the unmodified potential \eqref{eq:CR_potential}. This range depends on the model parameters and is illustrated in the left panel of Fig.~\ref{fig:background}, where $\phi_0$ can be calculated as
\begin{equation}
\frac{\phi_0}{M_{\mathrm{Pl}}}=\sqrt{\frac{2}{\beta}}\sin^{-1}\left\{e^{-N_*\beta}\sin\left[\frac{1}{2}\cos^{-1}\left(\frac{\beta-3}{\beta+3}\right)\right]\right\}.
\label{eq:phi_I}
\end{equation}
We note that $\phi_0$ is independent of $M$. Thus, the model has three parameters, namely $M$, $\beta$ and $N_*$. It is clear that $N_*$ and $M$ are degenerate since amplitude of scalar fluctuations can be changed by either a shift in the energy scale of the potential, or by moving the endpoint $\phi_0$ closer or further from $\phi=0$ where potential is flat, up to the value where $V=0$. 
In a similar way, it is necessary to define the field value $\phi_{\mathrm{p}}$ in which the scalar/tensor CMB pivot modes exit the horizon   
\begin{equation}
\frac{\phi_{\mathrm{p}}}{M_{\mathrm{Pl}}}=\sqrt{\frac{2}{\beta}}\sin^{-1}\left[e^{-55\beta}\sin\left(\sqrt{\frac{\beta}{2}}\frac{\phi_0}{M_{\mathrm{Pl}}}\right)\right],
\label{eq:phi_p}
\end{equation}
this is located 55 e-folds before the end of inflation. Using the equations of motion for the field \eqref{eq:back_eq_mov} and the Hubble scale \eqref{eq:Friedmann_eqns}, we scan the phase space in order to find the number of e-folds for every choice of initial conditions inside the interval $\phi/M_{\mathrm{Pl}}\in[-3;3]$ and $\dot{\phi}/M_{\mathrm{Pl}}^2\in[-3;3]$. Our results are shown in the right panel of Fig.~\ref{fig:background} for $M^2=2.0\times10^{-9}M_{\mathrm{Pl}}^2$, $N_*=55$ and $\beta=0.02$, where $\phi_0=3.38$ corresponds to our choice for $\beta$. A few phase space trajectories are also plotted in the same figure. A choice of initial conditions close to the attractor (with $\dot{\phi}$ small in this case) generates more expansion before reaching $\pm\phi_0$ due to a slow convergence to the attractor of the trajectories starting away from it. We take $\dot{\phi}=0$ as a suitable initial condition for the background field velocity that always reaches the attractor, and start numerical evolution of the background sufficiently far in the past for trajectory to settle to the attractor before considering fluctuations.

\section{Perturbations}
\label{sec:Perturbations}

\begin{figure*}[tbh]
\centering
\subfigure{
\includegraphics[width=.465\textwidth]{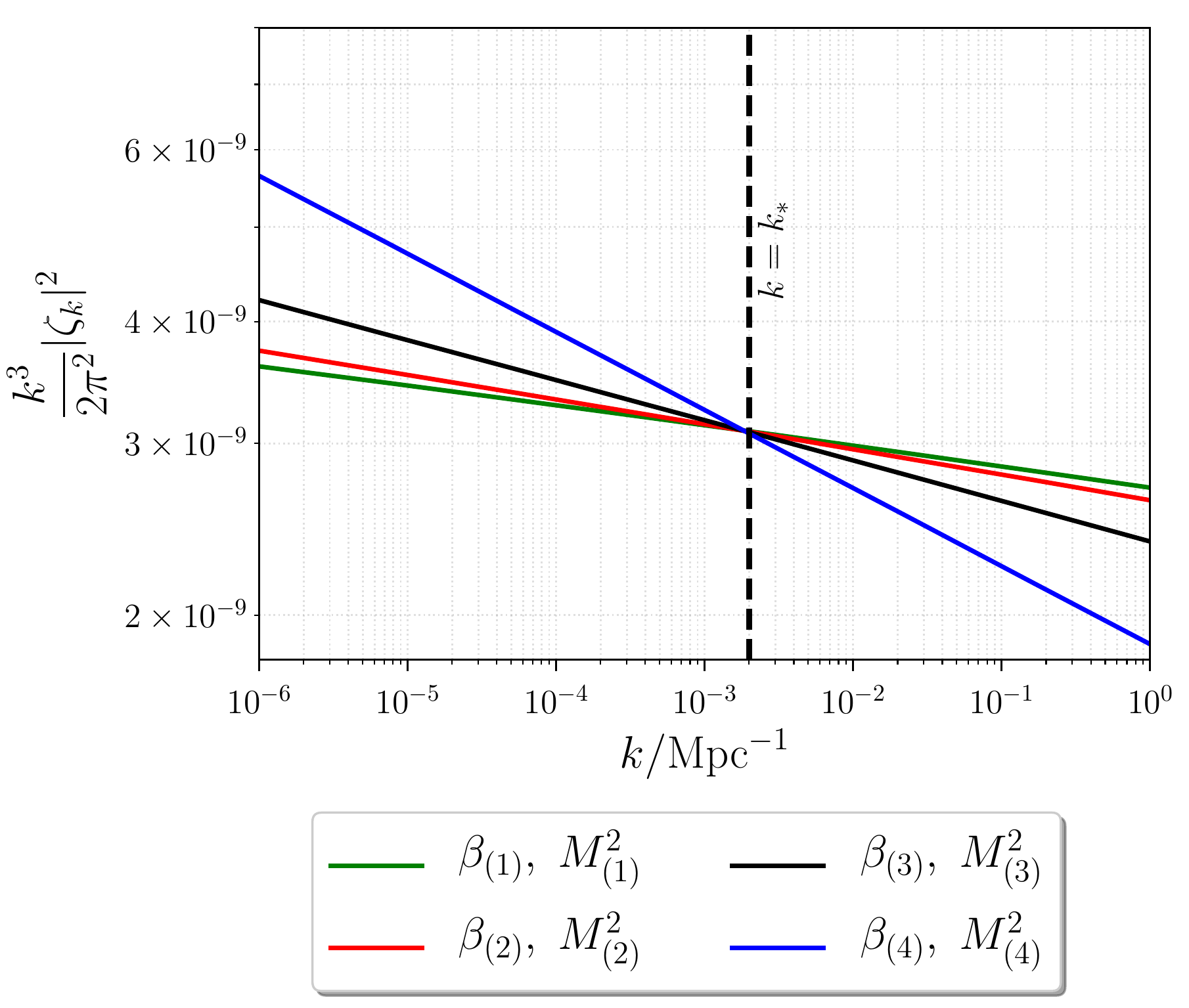}} \,
\subfigure{
\includegraphics[width=.465\textwidth]{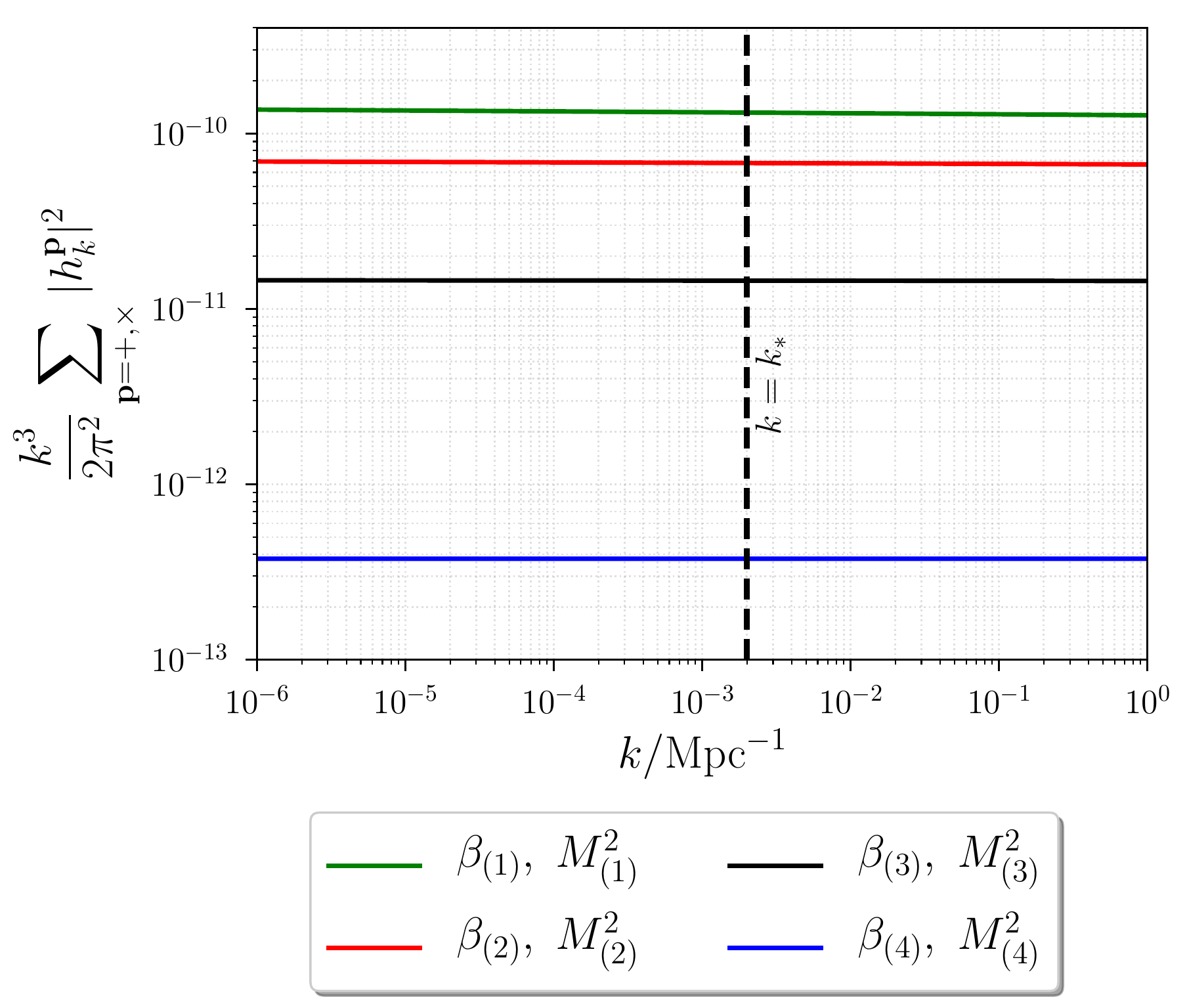}}
\caption{\label{fig:power_spectra} Left panel: Power spectrum of primordial curvature fluctuations using $k_*=2\times10^{-3} {\rm Mpc}^{-1}$ as a pivot scale at $N_*=0$. We considered the model parameters $\beta_{(1)}=0.005$, $M^2_{(1)}=10^{-9}M_{\mathrm{Pl}}^2$, $\beta_{(2)}=0.01$, $M^2_{(2)}=3.82\times10^{-10}M_{\mathrm{Pl}}^2$, $\beta_{(3)}=0.02$, $M^2_{(3)}=7.26\times10^{-11}M_{\mathrm{Pl}}^2$, $\beta_{(4)}=0.04$ and $M^2_{(4)}=1.86\times10^{-12}M_{\mathrm{Pl}}^2$. Right panel: Power spectrum of tensor perturbations for the same model parameters. None of the spectra show any features or running.} 
\end{figure*}

In this section, we recount the standard treatment of scalar and tensor perturbations required to obtain the power spectra of primordial fluctuations, and describe an extremely effective way to do so numerically. Let us first recall the expansion of the action \eqref{eq:action} up to second order in perturbations
\begin{eqnarray}
S_2&=\displaystyle{\frac{M_{\mathrm{Pl}}^2}{8}\int a^2 d\eta~d^3 x \left[h'_{ij}h'^{ij}-(\nabla_kh_{ij})(\nabla^kh^{ij})\right]}\nonumber\\
&+\displaystyle{\frac{1}{2}\int \left(\frac{\phi'^2}{H^2}\right) d\eta~d^3x \left[\zeta'^2-(\nabla_i\zeta)(\nabla^i\zeta)\right]},\label{eq:pert_action}
\end{eqnarray}
which is written in the gauge $\delta\phi=0$ and in conformal time $\eta$. Latin indices are raised and lowered by the Kronecker delta and primes denote derivatives w.r.t. $\eta$. In terms of the polarization modes $h_{ij}\equiv\sum_{\mathbf{p}=+,\times}h^{\mathbf{p}}\sigma^{\mathbf{p}}_{ij}$, the last expression is equivalent to
\begin{eqnarray}
S_2&=\displaystyle{\frac{M_{\mathrm{Pl}}^2}{8}\sum_{\mathbf{p}=+,\times}\int a^2 d\eta~d^3 x \left[\left(h^{\mathbf{p}'}\right)^2-(\nabla_kh^{\mathbf{p}})(\nabla^kh^{\mathbf{p}})\right]}\nonumber\\
&+\displaystyle{\frac{1}{2}\int \left(\frac{\phi'^2}{H^2}\right) d\eta~d^3x \left[\zeta'^2-(\nabla_i\zeta)(\nabla^i\zeta)\right]},
\end{eqnarray}
where we used the fact that $\mathrm{tr} \left(\sigma^+\right)^2=\mathrm{tr} \left(\sigma^{\times}\right)^2=1$ and $\mathrm{tr} \left(\sigma^+\sigma^{\times}\right)=0$. Now, we introduce the Mukhanov-Sasaki variables  $v\equiv \phi'/H\, \zeta$ and $v^{\mathbf{p}}\equiv aM_{\mathrm{Pl}}\,h^{\mathbf{p}}/2$ \cite{Mukhanov:1985rz, Sasaki:1986hm, Mukhanov:1988} to rewrite the action as
\begin{eqnarray}
&S_2=\displaystyle{\frac{1}{2}\sum_{\mathbf{p}=+,\times}\int d\eta~d^3 x \bigg[\left(v^{\mathbf{p}'}\right)^2-(\nabla_kh^{\mathbf{p}})(\nabla^kh^\mathbf{p})}\nonumber\\
&+\displaystyle{\frac{a''}{a}(v^{\mathbf{p}})^2\bigg]+\frac{1}{2}\int d\eta~d^3x \left[v'^2-(\nabla_iv)(\nabla^iv)+\frac{z''}{z}v^2\right]},\nonumber
\end{eqnarray}
where $z=\phi'/\sqrt{2}M_{\rm Pl}H$. Hence, the action for perturbations is now canonically normalized. In Fourier space, the equations of motion for fluctuations are given by
\begin{eqnarray}
&\displaystyle{v_k''+\left(k^2-\frac{z''}{z}\right)v_k=0,}\label{eq:scalar_pert}\\
&\displaystyle{\left(v^{\mathbf{p}}_k\right)''+\left(k^2-\frac{a''}{a}\right)v^{\mathbf{p}}_k=0,}\label{eq:tensor_pert}
\end{eqnarray}
where the Fourier transformed Mukhanov-Sasaki variables are $v_k\equiv \phi'/H\,\zeta_k$ and $v_k^{\mathbf{p}}\equiv aM_{\mathrm{Pl}}\,h_k^{\mathbf{p}}/2$. Both \eqref{eq:scalar_pert} and \eqref{eq:tensor_pert} have the form a harmonic oscillator with time-dependent frequency
\begin{equation}
\xi_k''+\omega^2_{\mathrm{eff}}(\eta)\xi_k=0.
\label{eq:TDHO}
\end{equation}
We will now apply a very simple trick which is incredibly effective for numerical evaluation of the perturbation spectra, which is the single-field version of the general method \citep{Ghersi:2016gee}. The fluctuation variable $\xi_k$ can be redefined in terms of real amplitude $L_k$ and phase $\Theta_k$ as $\xi_k\equiv L_k\exp\left(i\Theta_k\right)$. Substituting this ansatz into \eqref{eq:TDHO} splits the differential equation into real and imaginary parts
\begin{eqnarray}
&L_k''+\left[\omega_{\mathrm{eff}}^2(\eta)-(\Theta_k')^2\right]L_k=0, \label{eq:amplitude}\\
&\displaystyle{\Theta''_{k}+2\frac{L'_k}{L_k}\,\Theta_k'=0}, \label{eq:phase}
\end{eqnarray}
where the imaginary part \eqref{eq:phase} is separable and has a simple analytic solution $\Theta_k'(\eta)=\Theta_k'(\eta_0)L^2_k(\eta_0)/L^2_k(\eta)$. Once the phase is eliminated from \eqref{eq:amplitude}, we obtain 
\begin{equation}
L_k^{\prime \prime}+\left[\omega_{\mathrm{eff}}^2(\eta)-\omega_{\rm eff}^{ 2}(\eta_0) \frac{L^4_k(\eta_0)}{L^4_k(\eta)} \right]L_k=0, \label{eq:amplitude:only}
\end{equation}
where the Bunch-Davies vacuum deep inside horizon at $\eta=\eta_0$ sets $\Theta_k'(\eta_0)=\omega_{\mathrm{eff}}(\eta_0)$, $L_k=1/\sqrt{2\omega_{\mathrm{eff}}}$, and $L'_k(\eta_0)=0$ as initial conditions for mode evolution. The key observation is that the last term in \eqref{eq:amplitude:only} cancels the effective mode oscillation frequency, allowing numerical evolution to keep track of changes in amplitude only, with precision increasing deep inside the horizon where the vacuum state is more accurate. One no longer needs to resolve exponentially large physical oscillation scale $(k/a)$ inside horizon, and can use a time step merely a fraction of the Hubble scale to resolve the evolution of the amplitudes without compromising the precision of the evolution routine. The latter expression is also known as the Ermakov-Pinney equation \citep{Ermakov:1880, 10.2307/2032300, Kamenshchik:2005kf}. The implementation of this simple technique in equations \eqref{eq:scalar_pert} and \eqref{eq:tensor_pert} allowed us to calculate the evolution of the scalar curvature and the tensor fluctuations to extremely high precision, as shown in the two panels of Fig.~\ref{fig:pert_evol}. 

We can evaluate the power spectra of scalar and tensor perturbations once we compute the evolution of the scalar and tensor modes for relevant wavenumbers. As an illustration, we calculated the spectra for several sets of the model parameters choosing $k_*=2\times10^{-3} M_{\mathrm{pc}}^{-1}$ as a pivot scale in Fig.~\ref{fig:power_spectra}. The most striking fact to observe in both panels is the absence of any features or running in the spectra including $\beta=0.02$, as it was argued in \citep{Motohashi:2017aob}. This can be checked for any number of the modes used to produce each of the spectra. Thus, the estimations made in that paper about the shape of the spectrum, in analogy with the approximate treatment for natural inflation (see \citep{Freese:2014nla} for further details) are perfectly valid. Nevertheless, we can increase accuracy of the parameter estimation (especially away from the slow-roll regime) by calculating the power spectra directly for each of the model realizations. 

\begin{figure}[t]
\centering
\includegraphics[width=.43\textwidth]{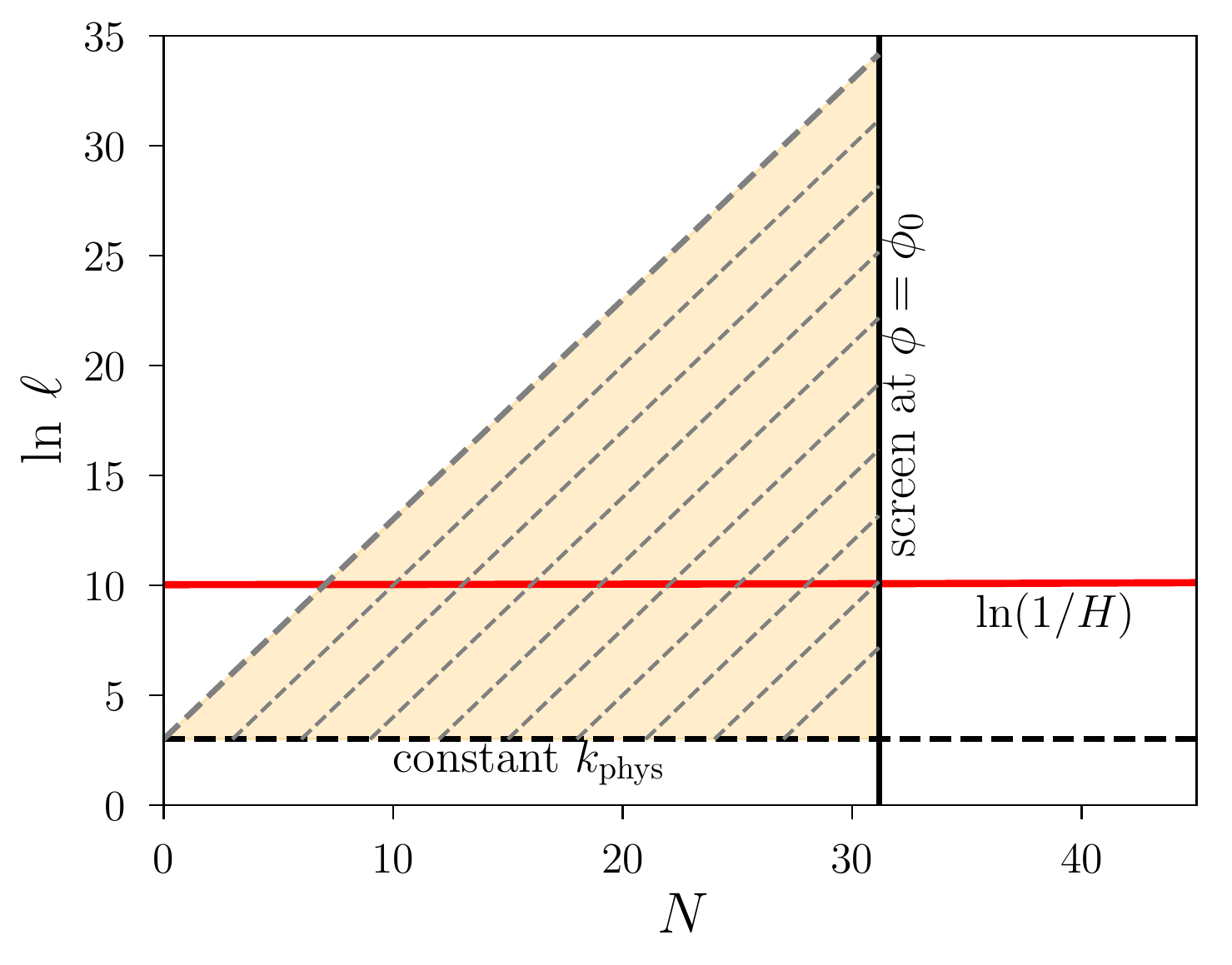}
\caption{\label{fig:evolscheme} Injection scheme for the scalar and tensor modes. Evolution begins at a surface of initial conditions deep inside the horizon, with modes injected at constant $k_{\mathrm{phys}}$.}
\end{figure}

To further improve the efficiency of our calculations in Fig.~\ref{fig:power_spectra}, we use the approximate time-translational symmetry of the Bunch-Davies vacuum deep inside the horizon, and only keep track of the physical wavelengths we are interested in. Scalar and tensor modes are evolved from a constant physical length scale $10^3$ times smaller than the horizon and then collected at moment of time when $\phi=\phi_0$ as shown in the mode injection scheme depicted in Fig~\ref{fig:evolscheme}. The length scale $1/H$ where the modes freeze out is plotted in red. Comoving modes evolve from $\lambda_{\mathrm{phys}}=10^{-3}/H$ across the physical length scale $\ell$ following the lines of constant comoving wavenumber (which have a slope of 1) until they reach the screen at $\phi=\phi_0$, where the mode amplitudes and the power spectra are evaluated. We can safely omit the evolution of the modes at physical scales shorter than injection point $k_{\mathrm{phys}}$ (below the orange triangle) as the vacuum is essentially stationary there.

\section{Planck Constraints} 
\label{sec:Observational_Constraints}

\begin{figure*}[t]
\centering
\subfigure{
\includegraphics[width=.45\textwidth]{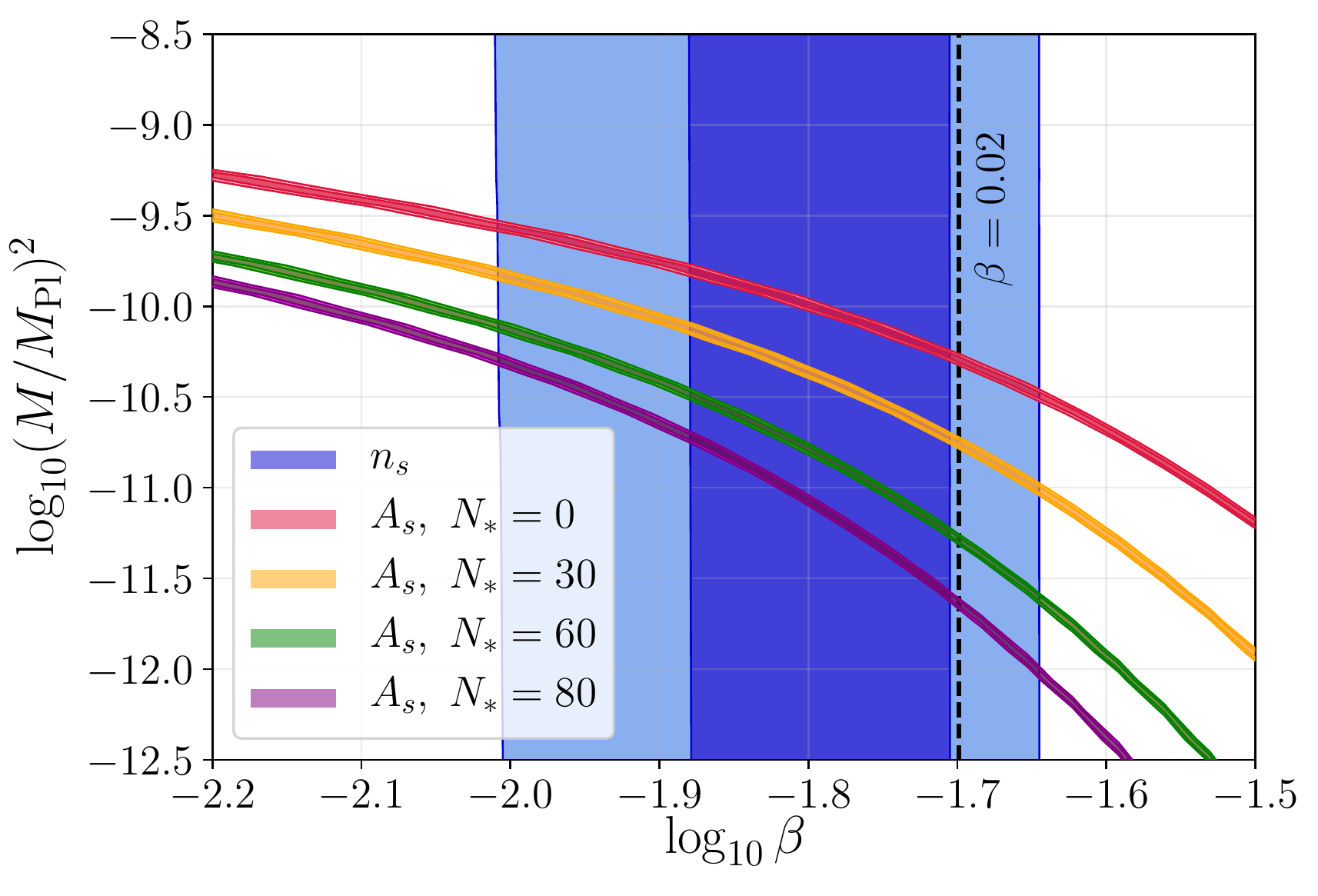}}\,
\subfigure{
\includegraphics[width=.45\textwidth]{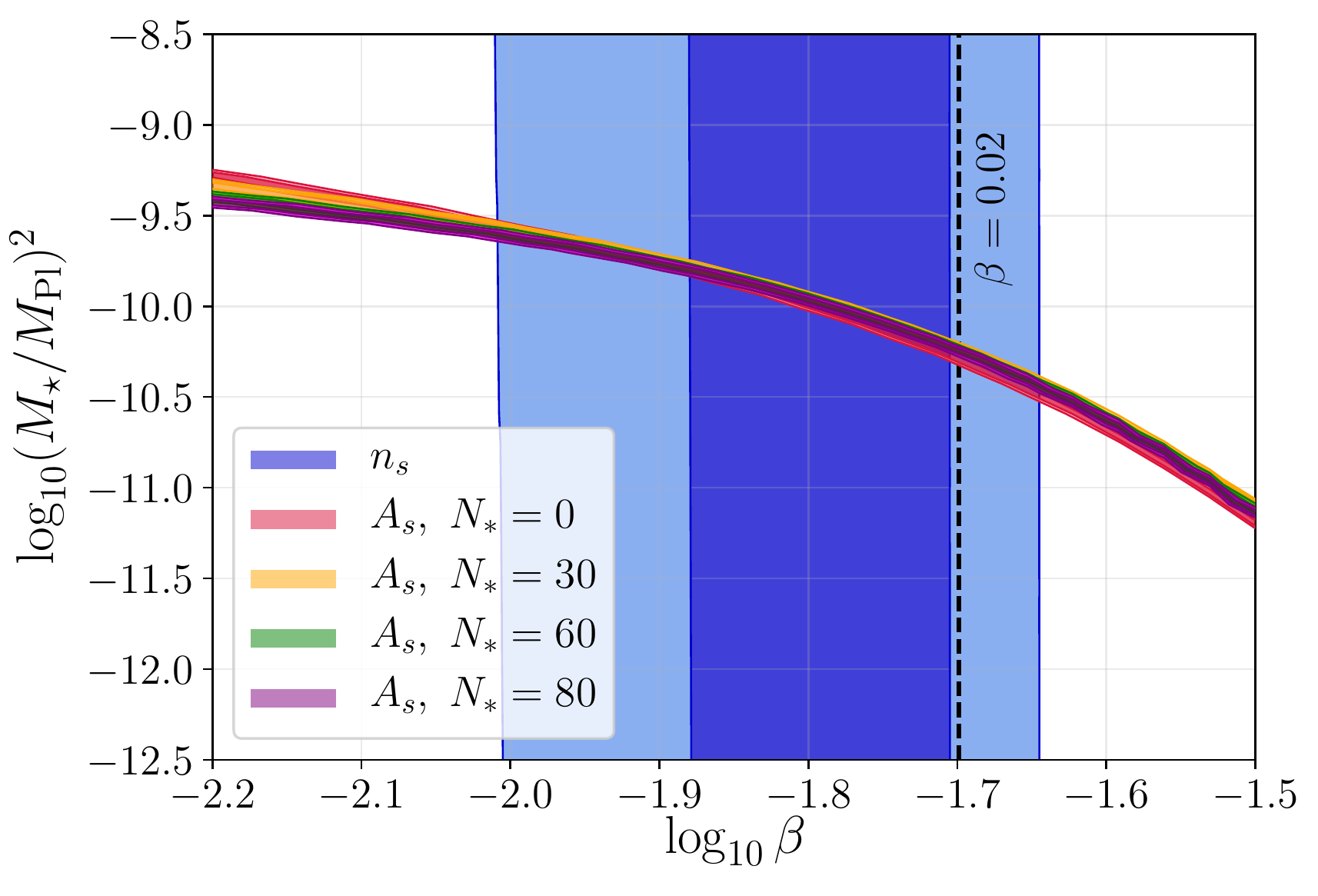}}
\caption{\label{fig:ModelParameters_Sampling_full} Scanning the parameters in the potential at $N_*=0, 30, 60, 80$. The blue regions show the 68\% C.L. (solid blue) and 95\%C.L. (shaded blue) regions for the marginalized posterior probability of the scalar spectral index $n_s$, while the regions for the marginalized posterior of the scalar primordial amplitude $A_s$ are plotted in different colors. On the left panel we can see the degeneracy of $N_*$ and $M^2$ with respect to the amplitude. The right panel shows that the degeneracy is resolved fairly well when we use $M_{\star}$ defined in Eq.~\eqref{eq:new_parameter} instead of $M$.  The black dashed lines in both panels represent the estimate of the optimal $\beta$ computed in \citep{Motohashi:2017aob}.}
\end{figure*}

In this section we use the CMB anisotropies measurements from the Planck satellite \citep{Adam:2015rua, Ade:2015xua, Ade:2015lrj} and their joint analysis with the BICEP2/Keck Array \citep{Ade:2015tva} to derive the constraints on the model parameters of the constant roll inflation.
On the left panel of Fig.~\ref{fig:ModelParameters_Sampling_full} we show the constraints on the parameters $M^2$ and $\beta$ from the Planck constraints on the scalar amplitude $A_s$ and spectral index $n_s$.
 The blue band corresponds to the region constrained by the scalar spectral index, $n_s$, while the other colored bands corresponds to the constrained regions from the scalar amplitude for different $N_*$ values. Here the degeneracy between the parameters $M$ and $N_*$ mentioned in Sect.\ref{sec:Model_Background} is evident: for a fixed value of $\beta$ one can raise or lower the value of $M$ and keep $A_s$ constant by adjusting $N_*$.  
To deal with this degeneracy one can proceed in two ways. Either we fix $N_*$ to some arbitrary value and we constrain $M$ for that choice, or we can combine the two variables $M$ and $N_*$ into one that parametrizes the degeneracy. We choose the latter and to find the combination of the two parameters we work in the following way. In the exact slow roll approximation we have, for $N_*$ adequately large,
\begin{equation}
A_s \propto \frac{H^2}{\epsilon} \simeq \frac{ M^2 \sinh^2 (\beta N_*)}{\beta} \sim \frac{M^2}{\beta}e^{2N_*\beta},
\end{equation}
where $H^2$ is the Hubble parameter a the end of inflation evaluated analytically in \citep{Motohashi:2017aob} and the slow-roll parameter $\epsilon$ is defined in \eqref{eqn:epsilon_slow_roll}.
Thus we can constrain the combination
\begin{equation}
\label{eq:new_parameter}
 M_{\star} \equiv M^2 \exp( 2N_* \beta)
 \end{equation}
 for which $A_s$ is constant for a fixed $\beta$. We evaluate the choice of this parameter graphically in the right panel of Fig.~\ref{fig:ModelParameters_Sampling_full}. We can see that for different $N_*$ values, the constrained area lies in the same range of $M_{\star}$.

As mentioned earlier, observational constraints on the constant-roll inflation parameters were already derived in \citep{Motohashi:2017aob} although by means of the slow-roll approximation. Here we improve on those earlier results as we are able to very accurately compute the power spectra of the scalar and tensor fluctuations using the evolution scheme presented in Sect.~\ref{sec:Perturbations}. Also, as discussed in Appendix~\ref{sec:AppendixSlowRoll}, deviations from the slow-roll conditions might be noticeable even in the best fit range of $\beta$ given the precision of the present day observational data.

\begin{figure}
\centering
\includegraphics[width=.45\textwidth]{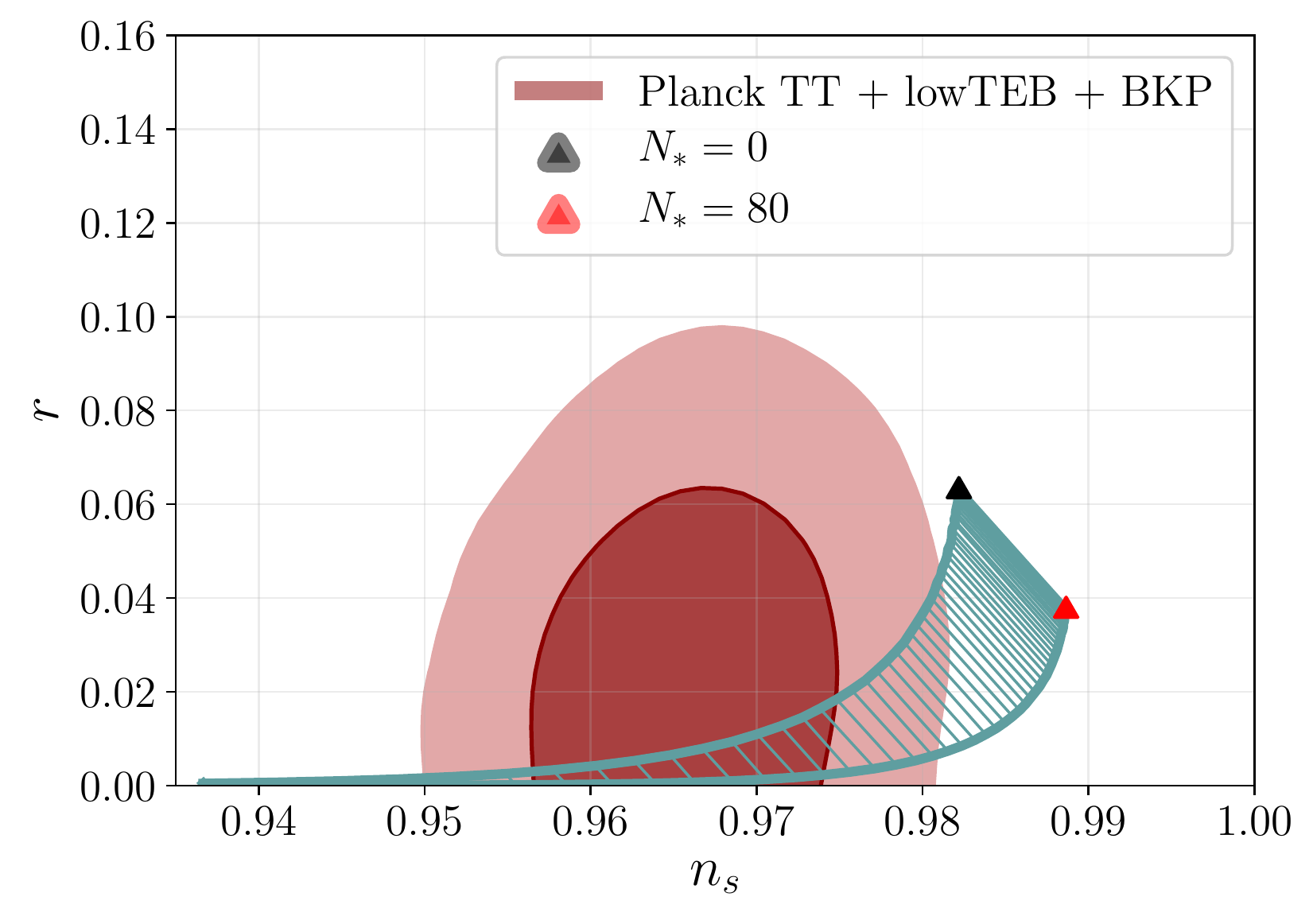}
\caption{\label{fig:TiltVSRatio} Constraints from joint Planck 2015/BKP likelihood on $n_s$ and $r$. The green line shows the values of $(n_s, r)$ from the parameter space probed in Figs.~\ref{fig:ModelParameters_Sampling_full} and \ref{fig:ModelParameters_Sampling}. For higher values of $N_*$, the model can cover most of the lower range of $r$.}
\end{figure}

\begin{figure*}[t]
\centering
\includegraphics[width=.7\textwidth]{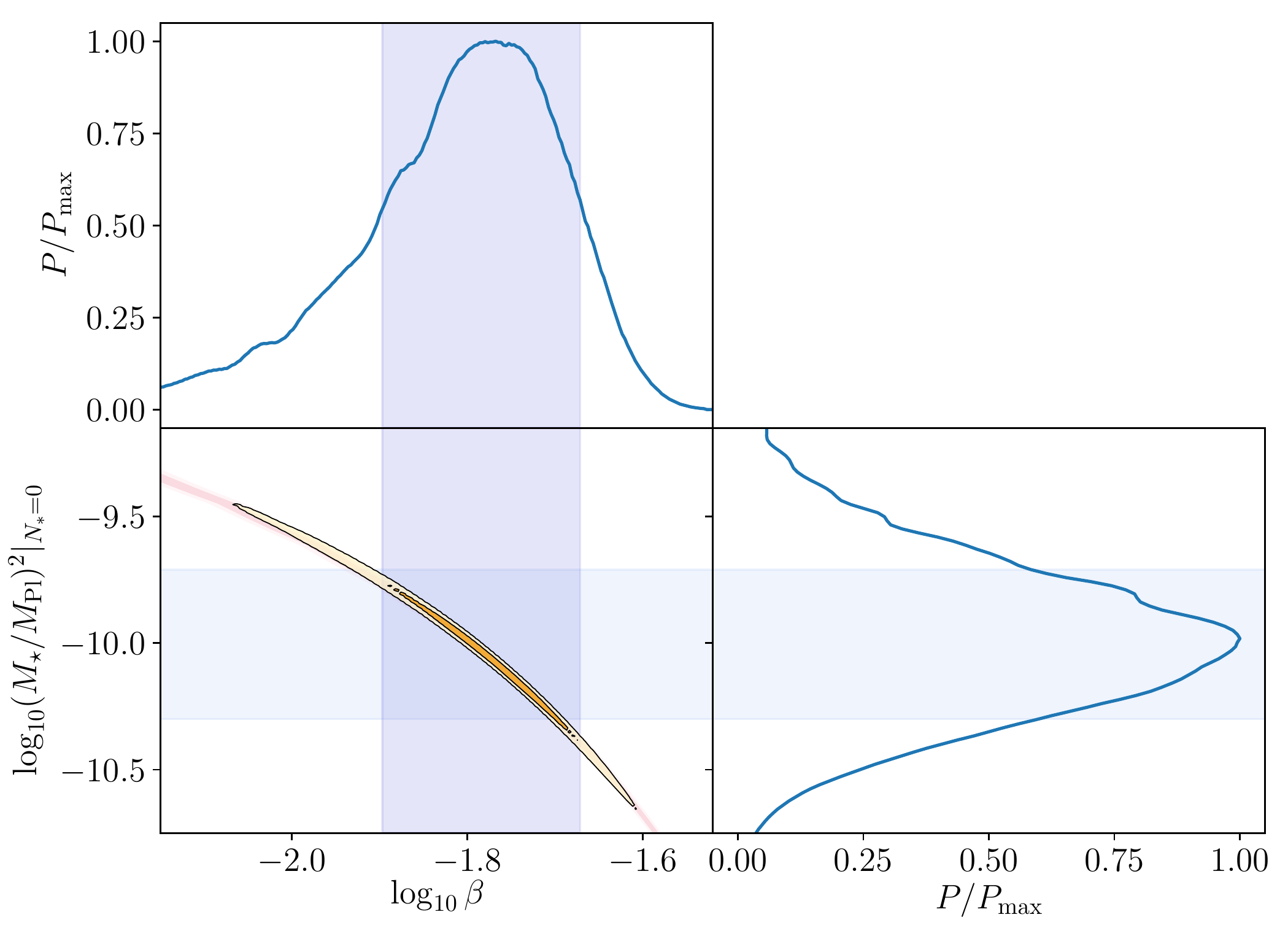}
\caption{\label{fig:ModelParameters_Sampling} Different projections of the joint probability distribution from Planck 2015 likelihood evaluated at $N_*=0$. Top panel: lateral projection of the joint distribution for $\beta$. Right panel: lateral projection of the joint distribution for $M_{\star}/M_{\rm Pl}|_{N_*=0}$. Left corner on the bottom: different regions of the joint posterior distribution are shown within the red stripe in orange at the 68\% C.L. (solid orange) and 95\% C.L. (shaded orange).The blue shaded regions correspond to the $1\sigma$ marginalized regions for the two parameters $\beta$ and $M_{\star}$. }
\end{figure*}

In a Bayesian framework the posterior probability for the model parameters is usually sampled through Markov Chain Monte Carlo (MCMC) engines such as CosmoMC \citep{Lewis:2002ah} or MontePython \citep{MontePython} coupled to a Boltzmann solver such as CAMB \citep{Lewis:1999bs} or CLASS \citep{CLASS}. The constraints on the model parameters are then derived  by marginalization of the posterior probability. 
In this work however, we derived the constraints on constant-roll inflation inflation by simply mapping the posterior probability on the parameters ($n_s$, $A_s$, $r$) to two of the constant-roll inflation parameters ($\beta$, $M^2$) at a fixed $N_*$. 
To do so, we evaluated the scalar and tensor power spectra on the logarithmic grid $\log_{10} \beta \in [-3,-1.5]$, $\log_{10} (M/M_{\rm Pl})^2 \in [-15,-7]$ for different values of $N_*=0,~30,~60,~80$, and computed the parameters $n_s$, $A_s$, $r$, $n_T$ for each sample. The absence of features in the power spectra, as shown in Fig.~\ref{fig:power_spectra}, allowed us to quickly obtain the scalar and tensor spectral indexes by a simple linear regression in more than 8000 different model realizations. 

The joint-posterior distribution over the parameters $\beta$ and the new parameter, $M_{\star}$, was then computed according to
\begin{equation}
\label{eq:PosteriorProabability}
\mathcal{P}\left[ \log_{10} \beta , \log_{10} \left(\frac{M_{\star}}{M_{\rm Pl}}\right)^2 \right] = \mathcal{P}(n_s, \ln A_s) J,
\end{equation}
where the Jacobian $J$ was computed numerically from the results
\begin{align}
n_s & = n_s \left[ \log_{10} \beta , \log_{10} \left( \frac{M_{\star}}{M_{\rm Pl}} \right)^2 \right],  \\
\ln A_s & = \ln A_s \left[ \log_{10}\beta ,  \log_{10}\left( \frac{M_{\star}}{M_{\rm Pl}} \right)^2 \right].
\end{align}
The posterior joint distribution $\mathcal{P} ( n_s , \ln A)$ was generated from the MCMC chains provided by the Planck collaboration\footnote{\url{https://pla.esac.esa.int/pla/}}\footnote{\url{https://wiki.cosmos.esa.int/planckpla2015/index.php/Cosmological_Parameters}}.
The results are shown in Fig.~\ref{fig:ModelParameters_Sampling}. The orange shaded regions represent the joint posterior probability $\mathcal{P}(\log_{10} \beta , \log_{10} (M_{\star} / M_{\rm Pl})^2)$, while the  blue shaded regions represent respectively the marginalized $1\sigma$ regions for the parameters $( \log_{10} \beta , \log_{10}(M_{\star} / M_{\rm Pl})^2)$. From the definition in \eqref{eq:new_parameter}, the constraints on the constant-roll inflation parameters are obtained by marginalizing the posterior probability \eqref{eq:PosteriorProabability} and are $\log_{10} \beta = -1.77^{+0.17}_{-0.35}$ and $\log_{10} (M_{\star}/M_{\rm {Pl}})^2 = -9.98_{-0.6}^{+0.7}$ at 95\% C.L.\\

In Fig.~\ref{fig:TiltVSRatio}, we illustrate the region of constant-roll model parameters we probed in the $r$ versus $n_s$ diagram overlaying the joint likelihood distribution provided by Planck 2015. The hatched region corresponds to the 
variation of $N_*$ spanning values from $0$ to $80$, while the $M$ and $\beta$ range as in Fig.~\ref{fig:ModelParameters_Sampling_full}. If one is willing to increase $N_*$ further (corresponding to hill-top inflation) very small values of $r$ can be achieved. Interestingly, the parameter region of the constant-roll inflation does not overlap with any of the existing regions constrained by other models shown in \citep{Ade:2015tva}, which makes constant-roll inflation a testable alternative for future observations.

\section{Discussions} 
\label{sec:Discussions}

In this paper, we provide constraints of the model parameters in constant-roll inflation, as proposed in \citep{Motohashi:2017aob,Motohashi:2014ppa}. These are not the only efforts regarding models with similar features, for instance see \citep{Tzirakis:2007bf, PhysRevD.52.5486, Anguelova:2017djf, Cai:2017bxr, Yi:2017mxs, Nojiri:2017qvx, Motohashi:2017vdc, Karam:2017rpw, Morse:2018kda} for further examples. Our numerical procedure is optimized for an efficient evaluation of the scalar and tensor power spectra of primordial fluctuations, and can scan more than 8000 different choices of model parameters in a reasonable time on minimal computing hardware. It does not require assuming the slow-roll approximation as it is based on the direct computation of the cosmological parameters ($n_s,r,A_s,n_T$) from the featureless power spectra shown in Fig.~\ref{fig:power_spectra}. The code passes numerous accuracy tests and long-time integration of the mode evolution confirms that there is no spurious evolution on super-horizon scales.

In order to provide tight constraints of the model parameters, we needed to address the degeneracy between $M$ and $N_*$. We found $M_{\star}$ defined in \eqref{eq:new_parameter} to be a good auxiliary parameter that leaves the spectra almost invariant under different choices of $N_*$ for any fixed value of $M_{\star}$. After using the CMB measurements from the Planck Collaboration \citep{Adam:2015rua, Ade:2015xua, Ade:2015lrj} and their joint likelihood with the BICEP2/Keck Array \citep{Ade:2015tva}, we estimated $\log_{10} \beta =-1.77^{+0.17}_{-0.35}$ and $\log_{10} (M_{\star}/M_{\rm {Pl}})^2 = -9.98_{-0.6}^{+0.7}$ at 95\% C.L. for $N_*=0$, as shown in Fig.~\ref{fig:ModelParameters_Sampling}. The constraints for $\beta$ are not significantly modified by any different choice of $N_*$, however, due to the parameter degeneracy the same cannot be said about the constraints for $M$.     
The parameter range on $r$ versus $n_s$ diagram covered by constant-roll inflation in Fig.~\ref{fig:TiltVSRatio} does not appear to overlap with any of the regions covered by the other inflationary models considered in \citep{Ade:2015xua, Ade:2015lrj}, making this model observationally interesting for the next generations of CMB experiments. 

\appendix

\section{Deviations of $n_s$ and $r$ from the slow-roll expressions}
\label{sec:AppendixSlowRoll}

\begin{figure*}[tbh]
\centering
\subfigure{
\includegraphics[width=.47\textwidth]{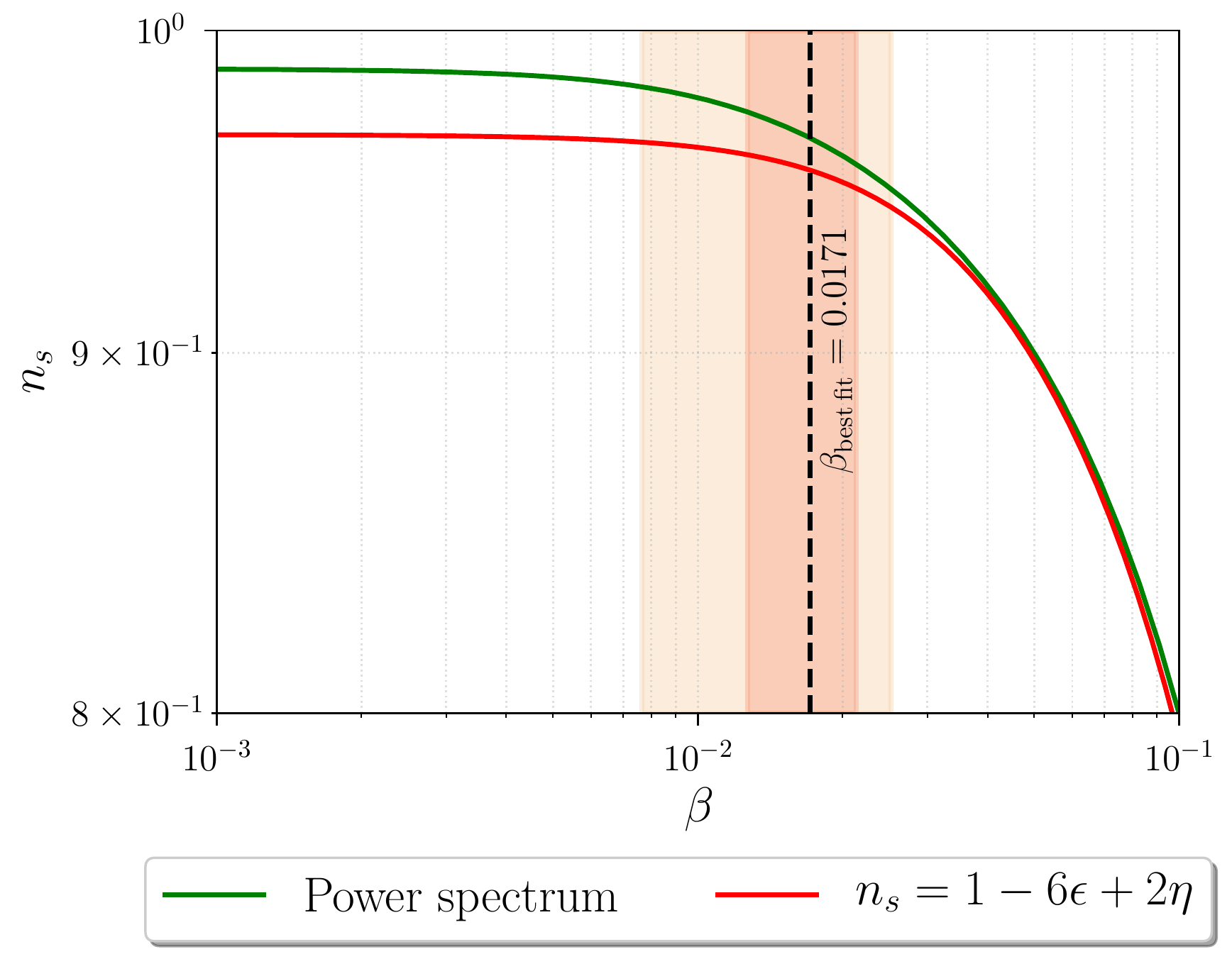}} \,
\subfigure{
\includegraphics[width=.47\textwidth]{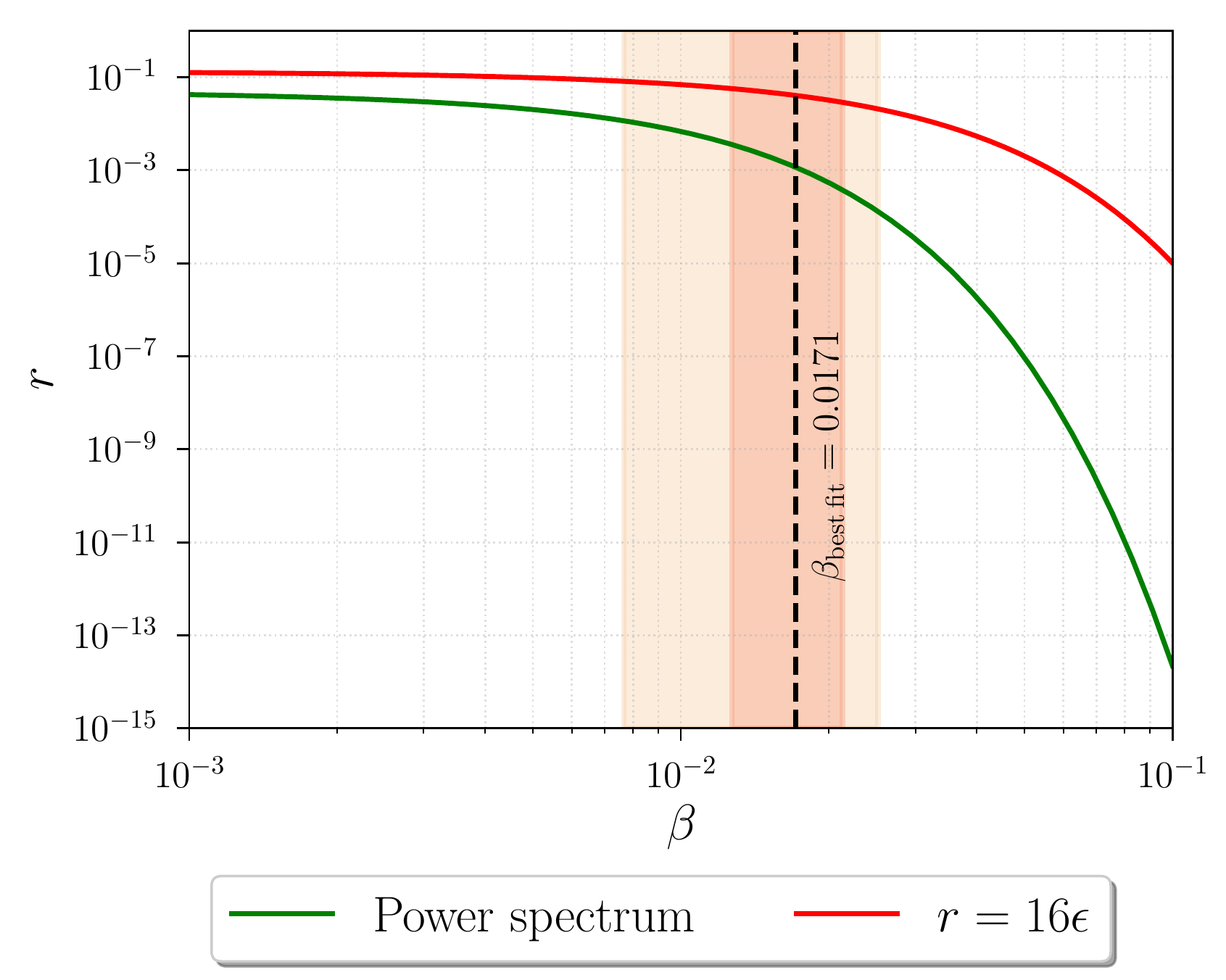}}
\caption{\label{fig:consistency} Left panel: Testing the consistency relation in \eqref{eq:spectral-index} as a function of $\beta$ after fixing $M^2=10^{-9}M_{\mathrm{Pl}}^2$. Surprisingly, the consistency relations work well for larger values of $\beta$. Right panel: Testing \eqref{eq:tensor-to-scalar} for the same value of $M^2$. Here the differences are very large (up to ten orders of magnitude) for larger values of $\beta$. Both consistency relations are compared with the direct calculation of $n_s$ and $r$ at $N_*=0$, right after evaluating the power spectra scalar and tensor perturbations. The shaded regions represent the ranges of $\beta$ within the 68\% and 95\% confidence levels.} 
\end{figure*}

In this appendix, we compare two of the consistency relations discussed in \citep{Liddle:1994dx, Gong:2001he}
\begin{eqnarray}
&r=16\epsilon,\label{eq:tensor-to-scalar}\\
&n_s=1-6\epsilon+2\eta,\label{eq:spectral-index}
\end{eqnarray} 
with the values of $n_s$ and $r$ obtained from the scalar and tensor power spectra, as shown in Fig.~\ref{fig:consistency}. These relations are the basis of the confrontation with Planck data made in \citep{Motohashi:2017aob}. The slow-roll parameters $\epsilon$ and $\eta$ appearing in \eqref{eq:tensor-to-scalar} and \eqref{eq:spectral-index} are calculated there to be
\begin{eqnarray}
\label{eqn:epsilon_slow_roll}
&\displaystyle{\epsilon=\frac{\beta(3+\beta)^2\sin^2(\sqrt{2\beta}\phi/M_{\mathrm{Pl}})}{\left[-3+\beta-(3+\beta)\cos(\sqrt{2\beta}\phi/M_{\mathrm{Pl}})\right]^2}},\label{eq:epsilon}\\
&\displaystyle{\eta=\frac{2\beta(3+\beta)\cos(\sqrt{2\beta}\phi/M_{\mathrm{Pl}})}{-3+\beta-(3+\beta)\cos(\sqrt{2\beta}\phi/M_{\mathrm{Pl}})}}.\label{eq:eta}
\end{eqnarray}
Fig.~\ref{fig:consistency} shows comparison of the slow-roll approximations \eqref{eq:tensor-to-scalar} and \eqref{eq:spectral-index} to the values inferred from the direct spectrum computation. Both slow-roll parameters are evaluated at $\phi=\phi_0$, which makes them independent of $M$. Our procedure allows us to evaluate both power spectra well beyond the slow-roll approximation, we show $\beta\in[0.001;0.1]$ which contains the reliability interval suggested in \citep{Motohashi:2017aob}. $M^2=10^{-9}M_{\mathrm{Pl}}^2$ is fixed as an arbitrary constant, which has no difference with its best-fit value at any other value of $N_*$. From Fig.~\ref{fig:consistency}, it is interesting to notice that the deviations from the slow-roll definition in \eqref{eq:tensor-to-scalar} are always non-negligible, especially in the range of large values of $\beta$. The magnitude of scalar perturbations is of $\mathcal{O}(1)$ at $\beta\geq 0.05$ and grows continuously as $\beta$ increases, implying that the perturbative approach is no longer valid in this regime for $\beta$. The opposite occurs with the deviations from \eqref{eq:spectral-index} as we can observe in the left panel. In either of these cases, it is beneficial to avoid the approximation derived from the slow-roll approximation in order to calculate $n_s$ and $r$ since (i) this imposes restrictions on the valid range of $\beta$ and (ii) as shown in Fig.~\ref{fig:consistency}, the calculation of both parameters from the power spectra shows noticeable deviations from the approximate expressions in \eqref{eq:tensor-to-scalar} and \eqref{eq:spectral-index} in both the original reliability range $\beta\in[0.005;0.025]$ and within the 95\% C.L. range, as depicted in both panels of the last figure. One possible cause is the spurious running of $n_s$ seen in the slow-roll consistency relations, which makes \eqref{eq:tensor-to-scalar} and \eqref{eq:spectral-index} depend slightly on the exact field value they are evaluated at. If one wants high precision, it is easier to just calculate the spectra directly rather than dealing with intricacies of the slow-roll expansion \citep{Karam:2017rpw} to get to the required expansion order.

\begin{acknowledgments}
We would like to thank Alexander Vikman for suggesting the idea of this project. We would also like to thank Bernard Carr and Jun'ichi Yokoyama for their valuable comments and discussions on the previous results. This project was partly funded by the Discovery Grants program of the Natural Sciences and Engineering Research Council of Canada and it was performed in part at the Perimeter Institute for Theoretical Physics. Research at the Perimeter Institute is supported by the Government of Canada through the Department of Innovation, Science and Economic Development Canada. JG is supported by the Billy Jones Graduate Scholarship, granted by the Physics Department at SFU. AZ is supported by the National Sciences and Engineering Research Council of Canada and the Bert Henry Graduate Scholarship.
\end{acknowledgments}

\bibliography{bibliography.bib}

\begin{thebibliography}{39}%
\makeatletter
\providecommand \@ifxundefined [1]{%
 \@ifx{#1\undefined}
}%
\providecommand \@ifnum [1]{%
 \ifnum #1\expandafter \@firstoftwo
 \else \expandafter \@secondoftwo
 \fi
}%
\providecommand \@ifx [1]{%
 \ifx #1\expandafter \@firstoftwo
 \else \expandafter \@secondoftwo
 \fi
}%
\providecommand \natexlab [1]{#1}%
\providecommand \enquote  [1]{``#1''}%
\providecommand \bibnamefont  [1]{#1}%
\providecommand \bibfnamefont [1]{#1}%
\providecommand \citenamefont [1]{#1}%
\providecommand \href@noop [0]{\@secondoftwo}%
\providecommand \href [0]{\begingroup \@sanitize@url \@href}%
\providecommand \@href[1]{\@@startlink{#1}\@@href}%
\providecommand \@@href[1]{\endgroup#1\@@endlink}%
\providecommand \@sanitize@url [0]{\catcode `\\12\catcode `\$12\catcode
  `\&12\catcode `\#12\catcode `\^12\catcode `\_12\catcode `\%12\relax}%
\providecommand \@@startlink[1]{}%
\providecommand \@@endlink[0]{}%
\providecommand \url  [0]{\begingroup\@sanitize@url \@url }%
\providecommand \@url [1]{\endgroup\@href {#1}{\urlprefix }}%
\providecommand \urlprefix  [0]{URL }%
\providecommand \Eprint [0]{\href }%
\providecommand \doibase [0]{http://dx.doi.org/}%
\providecommand \selectlanguage [0]{\@gobble}%
\providecommand \bibinfo  [0]{\@secondoftwo}%
\providecommand \bibfield  [0]{\@secondoftwo}%
\providecommand \translation [1]{[#1]}%
\providecommand \BibitemOpen [0]{}%
\providecommand \bibitemStop [0]{}%
\providecommand \bibitemNoStop [0]{.\EOS\space}%
\providecommand \EOS [0]{\spacefactor3000\relax}%
\providecommand \BibitemShut  [1]{\csname bibitem#1\endcsname}%
\let\auto@bib@innerbib\@empty
\bibitem [{\citenamefont {Starobinsky}(1980)}]{Starobinsky:1980te}%
  \BibitemOpen
  \bibfield  {author} {\bibinfo {author} {\bibfnamefont {A.~A.}\ \bibnamefont
  {Starobinsky}},\ }\href {\doibase 10.1016/0370-2693(80)90670-X} {\bibfield
  {journal} {\bibinfo  {journal} {Phys. Lett.}\ }\textbf {\bibinfo {volume}
  {B91}},\ \bibinfo {pages} {99} (\bibinfo {year} {1980})}\BibitemShut
  {NoStop}%
\bibitem [{\citenamefont {{Sato}}(1981)}]{Sato:1981}%
  \BibitemOpen
  \bibfield  {author} {\bibinfo {author} {\bibfnamefont {K.}~\bibnamefont
  {{Sato}}},\ }\href {\doibase 10.1093/mnras/195.3.467} {\bibfield  {journal}
  {\bibinfo  {journal} {Mon. Not. Roy. Astron. Soc.}\ }\textbf {\bibinfo
  {volume} {195}},\ \bibinfo {pages} {467} (\bibinfo {year}
  {1981})}\BibitemShut {NoStop}%
\bibitem [{\citenamefont {Guth}(1981)}]{PhysRevD.23.347}%
  \BibitemOpen
  \bibfield  {author} {\bibinfo {author} {\bibfnamefont {A.~H.}\ \bibnamefont
  {Guth}},\ }\href {\doibase 10.1103/PhysRevD.23.347} {\bibfield  {journal}
  {\bibinfo  {journal} {Phys. Rev. D}\ }\textbf {\bibinfo {volume} {23}},\
  \bibinfo {pages} {347} (\bibinfo {year} {1981})}\BibitemShut {NoStop}%
\bibitem [{\citenamefont {{Linde}}(1982)}]{Linde:1982}%
  \BibitemOpen
  \bibfield  {author} {\bibinfo {author} {\bibfnamefont {A.~D.}\ \bibnamefont
  {{Linde}}},\ }\href {\doibase 10.1016/0370-2693(82)91219-9} {\bibfield
  {journal} {\bibinfo  {journal} {Phys. Lett. B}\ }\textbf {\bibinfo {volume}
  {108}},\ \bibinfo {pages} {389} (\bibinfo {year} {1982})}\BibitemShut
  {NoStop}%
\bibitem [{\citenamefont {Albrecht}\ and\ \citenamefont
  {Steinhardt}(1982)}]{Albrecht:1982}%
  \BibitemOpen
  \bibfield  {author} {\bibinfo {author} {\bibfnamefont {A.}~\bibnamefont
  {Albrecht}}\ and\ \bibinfo {author} {\bibfnamefont {P.~J.}\ \bibnamefont
  {Steinhardt}},\ }\href {\doibase 10.1103/PhysRevLett.48.1220} {\bibfield
  {journal} {\bibinfo  {journal} {Phys. Rev. Lett.}\ }\textbf {\bibinfo
  {volume} {48}},\ \bibinfo {pages} {1220} (\bibinfo {year}
  {1982})}\BibitemShut {NoStop}%
\bibitem [{\citenamefont {Motohashi}\ \emph {et~al.}(2015)\citenamefont
  {Motohashi}, \citenamefont {Starobinsky},\ and\ \citenamefont
  {Yokoyama}}]{Motohashi:2014ppa}%
  \BibitemOpen
  \bibfield  {author} {\bibinfo {author} {\bibfnamefont {H.}~\bibnamefont
  {Motohashi}}, \bibinfo {author} {\bibfnamefont {A.~A.}\ \bibnamefont
  {Starobinsky}}, \ and\ \bibinfo {author} {\bibfnamefont {J.}~\bibnamefont
  {Yokoyama}},\ }\href {\doibase 10.1088/1475-7516/2015/09/018} {\bibfield
  {journal} {\bibinfo  {journal} {JCAP}\ }\textbf {\bibinfo {volume} {1509}},\
  \bibinfo {pages} {018} (\bibinfo {year} {2015})},\ \Eprint
  {http://arxiv.org/abs/1411.5021} {arXiv:1411.5021 [astro-ph.CO]} \BibitemShut
  {NoStop}%
\bibitem [{\citenamefont {Motohashi}\ and\ \citenamefont
  {Starobinsky}(2017{\natexlab{a}})}]{Motohashi:2017aob}%
  \BibitemOpen
  \bibfield  {author} {\bibinfo {author} {\bibfnamefont {H.}~\bibnamefont
  {Motohashi}}\ and\ \bibinfo {author} {\bibfnamefont {A.~A.}\ \bibnamefont
  {Starobinsky}},\ }\href {\doibase 10.1209/0295-5075/117/39001} {\bibfield
  {journal} {\bibinfo  {journal} {EPL}\ }\textbf {\bibinfo {volume} {117}},\
  \bibinfo {pages} {39001} (\bibinfo {year} {2017}{\natexlab{a}})},\ \Eprint
  {http://arxiv.org/abs/1702.05847} {arXiv:1702.05847 [astro-ph.CO]}
  \BibitemShut {NoStop}%
\bibitem [{\citenamefont {Freese}\ \emph {et~al.}(1990)\citenamefont {Freese},
  \citenamefont {Frieman},\ and\ \citenamefont {Olinto}}]{PhysRevLett.65.3233}%
  \BibitemOpen
  \bibfield  {author} {\bibinfo {author} {\bibfnamefont {K.}~\bibnamefont
  {Freese}}, \bibinfo {author} {\bibfnamefont {J.~A.}\ \bibnamefont {Frieman}},
  \ and\ \bibinfo {author} {\bibfnamefont {A.~V.}\ \bibnamefont {Olinto}},\
  }\href {\doibase 10.1103/PhysRevLett.65.3233} {\bibfield  {journal} {\bibinfo
   {journal} {Phys. Rev. Lett.}\ }\textbf {\bibinfo {volume} {65}},\ \bibinfo
  {pages} {3233} (\bibinfo {year} {1990})}\BibitemShut {NoStop}%
\bibitem [{\citenamefont {Adam}\ \emph {et~al.}(2016)\citenamefont {Adam} \emph
  {et~al.}}]{Adam:2015rua}%
  \BibitemOpen
  \bibfield  {author} {\bibinfo {author} {\bibfnamefont {R.}~\bibnamefont
  {Adam}} \emph {et~al.} (\bibinfo {collaboration} {Planck}),\ }\href {\doibase
  10.1051/0004-6361/201527101} {\bibfield  {journal} {\bibinfo  {journal}
  {Astron. Astrophys.}\ }\textbf {\bibinfo {volume} {594}},\ \bibinfo {pages}
  {A1} (\bibinfo {year} {2016})},\ \Eprint {http://arxiv.org/abs/1502.01582}
  {arXiv:1502.01582 [astro-ph.CO]} \BibitemShut {NoStop}%
\bibitem [{\citenamefont {Ade}\ \emph {et~al.}(2016{\natexlab{a}})\citenamefont
  {Ade} \emph {et~al.}}]{Ade:2015xua}%
  \BibitemOpen
  \bibfield  {author} {\bibinfo {author} {\bibfnamefont {P.~A.~R.}\
  \bibnamefont {Ade}} \emph {et~al.} (\bibinfo {collaboration} {Planck}),\
  }\href {\doibase 10.1051/0004-6361/201525830} {\bibfield  {journal} {\bibinfo
   {journal} {Astron. Astrophys.}\ }\textbf {\bibinfo {volume} {594}},\
  \bibinfo {pages} {A13} (\bibinfo {year} {2016}{\natexlab{a}})},\ \Eprint
  {http://arxiv.org/abs/1502.01589} {arXiv:1502.01589 [astro-ph.CO]}
  \BibitemShut {NoStop}%
\bibitem [{\citenamefont {Ade}\ \emph {et~al.}(2016{\natexlab{b}})\citenamefont
  {Ade} \emph {et~al.}}]{Ade:2015lrj}%
  \BibitemOpen
  \bibfield  {author} {\bibinfo {author} {\bibfnamefont {P.~A.~R.}\
  \bibnamefont {Ade}} \emph {et~al.} (\bibinfo {collaboration} {Planck}),\
  }\href {\doibase 10.1051/0004-6361/201525898} {\bibfield  {journal} {\bibinfo
   {journal} {Astron. Astrophys.}\ }\textbf {\bibinfo {volume} {594}},\
  \bibinfo {pages} {A20} (\bibinfo {year} {2016}{\natexlab{b}})},\ \Eprint
  {http://arxiv.org/abs/1502.02114} {arXiv:1502.02114 [astro-ph.CO]}
  \BibitemShut {NoStop}%
\bibitem [{\citenamefont {Ade}\ \emph {et~al.}(2015)\citenamefont {Ade} \emph
  {et~al.}}]{Ade:2015tva}%
  \BibitemOpen
  \bibfield  {author} {\bibinfo {author} {\bibfnamefont {P.~A.~R.}\
  \bibnamefont {Ade}} \emph {et~al.} (\bibinfo {collaboration} {BICEP2,
  Planck}),\ }\href {\doibase 10.1103/PhysRevLett.114.101301} {\bibfield
  {journal} {\bibinfo  {journal} {Phys. Rev. Lett.}\ }\textbf {\bibinfo
  {volume} {114}},\ \bibinfo {pages} {101301} (\bibinfo {year} {2015})},\
  \Eprint {http://arxiv.org/abs/1502.00612} {arXiv:1502.00612 [astro-ph.CO]}
  \BibitemShut {NoStop}%
\bibitem [{\citenamefont {Liddle}\ \emph {et~al.}(1994)\citenamefont {Liddle},
  \citenamefont {Parsons},\ and\ \citenamefont {Barrow}}]{Liddle:1994dx}%
  \BibitemOpen
  \bibfield  {author} {\bibinfo {author} {\bibfnamefont {A.~R.}\ \bibnamefont
  {Liddle}}, \bibinfo {author} {\bibfnamefont {P.}~\bibnamefont {Parsons}}, \
  and\ \bibinfo {author} {\bibfnamefont {J.~D.}\ \bibnamefont {Barrow}},\
  }\href {\doibase 10.1103/PhysRevD.50.7222} {\bibfield  {journal} {\bibinfo
  {journal} {Phys. Rev.}\ }\textbf {\bibinfo {volume} {D50}},\ \bibinfo {pages}
  {7222} (\bibinfo {year} {1994})},\ \Eprint
  {http://arxiv.org/abs/astro-ph/9408015} {arXiv:astro-ph/9408015 [astro-ph]}
  \BibitemShut {NoStop}%
\bibitem [{\citenamefont {Gong}\ and\ \citenamefont
  {Stewart}(2001)}]{Gong:2001he}%
  \BibitemOpen
  \bibfield  {author} {\bibinfo {author} {\bibfnamefont {J.-O.}\ \bibnamefont
  {Gong}}\ and\ \bibinfo {author} {\bibfnamefont {E.~D.}\ \bibnamefont
  {Stewart}},\ }\href {\doibase 10.1016/S0370-2693(01)00616-5} {\bibfield
  {journal} {\bibinfo  {journal} {Phys. Lett.}\ }\textbf {\bibinfo {volume}
  {B510}},\ \bibinfo {pages} {1} (\bibinfo {year} {2001})},\ \Eprint
  {http://arxiv.org/abs/astro-ph/0101225} {arXiv:astro-ph/0101225 [astro-ph]}
  \BibitemShut {NoStop}%
\bibitem [{\citenamefont {Ghersi}\ and\ \citenamefont
  {Frolov}(2017)}]{Ghersi:2016gee}%
  \BibitemOpen
  \bibfield  {author} {\bibinfo {author} {\bibfnamefont {J.~T.~G.}\
  \bibnamefont {Ghersi}}\ and\ \bibinfo {author} {\bibfnamefont {A.~V.}\
  \bibnamefont {Frolov}},\ }\href {\doibase 10.1088/1475-7516/2017/05/047}
  {\bibfield  {journal} {\bibinfo  {journal} {JCAP}\ }\textbf {\bibinfo
  {volume} {1705}},\ \bibinfo {pages} {047} (\bibinfo {year} {2017})},\ \Eprint
  {http://arxiv.org/abs/1609.04770} {arXiv:1609.04770 [astro-ph.CO]}
  \BibitemShut {NoStop}%
\bibitem [{\citenamefont {Akrami}\ \emph
  {et~al.}(2018{\natexlab{a}})\citenamefont {Akrami} \emph
  {et~al.}}]{Akrami:2018vks}%
  \BibitemOpen
  \bibfield  {author} {\bibinfo {author} {\bibfnamefont {Y.}~\bibnamefont
  {Akrami}} \emph {et~al.} (\bibinfo {collaboration} {Planck}),\ }\href@noop {}
  {\  (\bibinfo {year} {2018}{\natexlab{a}})},\ \Eprint
  {http://arxiv.org/abs/1807.06205} {arXiv:1807.06205 [astro-ph.CO]}
  \BibitemShut {NoStop}%
\bibitem [{\citenamefont {Aghanim}\ \emph {et~al.}(2018)\citenamefont {Aghanim}
  \emph {et~al.}}]{Aghanim:2018eyx}%
  \BibitemOpen
  \bibfield  {author} {\bibinfo {author} {\bibfnamefont {N.}~\bibnamefont
  {Aghanim}} \emph {et~al.} (\bibinfo {collaboration} {Planck}),\ }\href@noop
  {} {\  (\bibinfo {year} {2018})},\ \Eprint {http://arxiv.org/abs/1807.06209}
  {arXiv:1807.06209 [astro-ph.CO]} \BibitemShut {NoStop}%
\bibitem [{\citenamefont {Akrami}\ \emph
  {et~al.}(2018{\natexlab{b}})\citenamefont {Akrami} \emph
  {et~al.}}]{Akrami:2018odb}%
  \BibitemOpen
  \bibfield  {author} {\bibinfo {author} {\bibfnamefont {Y.}~\bibnamefont
  {Akrami}} \emph {et~al.} (\bibinfo {collaboration} {Planck}),\ }\href@noop {}
  {\  (\bibinfo {year} {2018}{\natexlab{b}})},\ \Eprint
  {http://arxiv.org/abs/1807.06211} {arXiv:1807.06211 [astro-ph.CO]}
  \BibitemShut {NoStop}%
\bibitem [{\citenamefont {Martin}\ \emph {et~al.}(2013)\citenamefont {Martin},
  \citenamefont {Motohashi},\ and\ \citenamefont {Suyama}}]{Martin:2012pe}%
  \BibitemOpen
  \bibfield  {author} {\bibinfo {author} {\bibfnamefont {J.}~\bibnamefont
  {Martin}}, \bibinfo {author} {\bibfnamefont {H.}~\bibnamefont {Motohashi}}, \
  and\ \bibinfo {author} {\bibfnamefont {T.}~\bibnamefont {Suyama}},\ }\href
  {\doibase 10.1103/PhysRevD.87.023514} {\bibfield  {journal} {\bibinfo
  {journal} {Phys. Rev.}\ }\textbf {\bibinfo {volume} {D87}},\ \bibinfo {pages}
  {023514} (\bibinfo {year} {2013})},\ \Eprint {http://arxiv.org/abs/1211.0083}
  {arXiv:1211.0083 [astro-ph.CO]} \BibitemShut {NoStop}%
\bibitem [{\citenamefont {Mukhanov}(1985)}]{Mukhanov:1985rz}%
  \BibitemOpen
  \bibfield  {author} {\bibinfo {author} {\bibfnamefont {V.~F.}\ \bibnamefont
  {Mukhanov}},\ }\href@noop {} {\bibfield  {journal} {\bibinfo  {journal} {JETP
  Lett.}\ }\textbf {\bibinfo {volume} {41}},\ \bibinfo {pages} {493} (\bibinfo
  {year} {1985})},\ \bibinfo {note} {[Pisma Zh. Eksp. Teor.
  Fiz.41,402(1985)]}\BibitemShut {NoStop}%
\bibitem [{\citenamefont {Sasaki}(1986)}]{Sasaki:1986hm}%
  \BibitemOpen
  \bibfield  {author} {\bibinfo {author} {\bibfnamefont {M.}~\bibnamefont
  {Sasaki}},\ }\href {\doibase 10.1143/PTP.76.1036} {\bibfield  {journal}
  {\bibinfo  {journal} {Prog. Theor. Phys.}\ }\textbf {\bibinfo {volume}
  {76}},\ \bibinfo {pages} {1036} (\bibinfo {year} {1986})}\BibitemShut
  {NoStop}%
\bibitem [{\citenamefont {Mukhanov}(1988)}]{Mukhanov:1988}%
  \BibitemOpen
  \bibfield  {author} {\bibinfo {author} {\bibfnamefont {V.~F.}\ \bibnamefont
  {Mukhanov}},\ }\href@noop {} {\bibfield  {journal} {\bibinfo  {journal} {Zh.
  Eksp. Teor. Fiz.}\ }\textbf {\bibinfo {volume} {94}},\ \bibinfo {pages} {1}
  (\bibinfo {year} {1988})}\BibitemShut {NoStop}%
\bibitem [{\citenamefont {Ermakov}(1880)}]{Ermakov:1880}%
  \BibitemOpen
  \bibfield  {author} {\bibinfo {author} {\bibfnamefont {V.~P.}\ \bibnamefont
  {Ermakov}},\ }\href@noop {} {\bibfield  {journal} {\bibinfo  {journal} {Univ.
  Izv. Kiev}\ }\textbf {\bibinfo {volume} {III}} (\bibinfo {year}
  {1880})}\BibitemShut {NoStop}%
\bibitem [{\citenamefont {Pinney}(1950)}]{10.2307/2032300}%
  \BibitemOpen
  \bibfield  {author} {\bibinfo {author} {\bibfnamefont {E.}~\bibnamefont
  {Pinney}},\ }\href {http://www.jstor.org/stable/2032300} {\bibfield
  {journal} {\bibinfo  {journal} {Proceedings of the American Mathematical
  Society}\ }\textbf {\bibinfo {volume} {1}},\ \bibinfo {pages} {681} (\bibinfo
  {year} {1950})}\BibitemShut {NoStop}%
\bibitem [{\citenamefont {Kamenshchik}\ and\ \citenamefont
  {Venturi}(2009)}]{Kamenshchik:2005kf}%
  \BibitemOpen
  \bibfield  {author} {\bibinfo {author} {\bibfnamefont {A.}~\bibnamefont
  {Kamenshchik}}\ and\ \bibinfo {author} {\bibfnamefont {G.}~\bibnamefont
  {Venturi}},\ }\href {\doibase 10.1007/s11182-010-9375-4} {\bibfield
  {journal} {\bibinfo  {journal} {Russ. Phys. J.}\ }\textbf {\bibinfo {volume}
  {52}},\ \bibinfo {pages} {1339} (\bibinfo {year} {2009})},\ \Eprint
  {http://arxiv.org/abs/math-ph/0506017} {arXiv:math-ph/0506017 [math-ph]}
  \BibitemShut {NoStop}%
\bibitem [{\citenamefont {Freese}\ and\ \citenamefont
  {Kinney}(2015)}]{Freese:2014nla}%
  \BibitemOpen
  \bibfield  {author} {\bibinfo {author} {\bibfnamefont {K.}~\bibnamefont
  {Freese}}\ and\ \bibinfo {author} {\bibfnamefont {W.~H.}\ \bibnamefont
  {Kinney}},\ }\href {\doibase 10.1088/1475-7516/2015/03/044} {\bibfield
  {journal} {\bibinfo  {journal} {JCAP}\ }\textbf {\bibinfo {volume} {1503}},\
  \bibinfo {pages} {044} (\bibinfo {year} {2015})},\ \Eprint
  {http://arxiv.org/abs/1403.5277} {arXiv:1403.5277 [astro-ph.CO]} \BibitemShut
  {NoStop}%
\bibitem [{\citenamefont {Lewis}\ and\ \citenamefont
  {Bridle}(2002)}]{Lewis:2002ah}%
  \BibitemOpen
  \bibfield  {author} {\bibinfo {author} {\bibfnamefont {A.}~\bibnamefont
  {Lewis}}\ and\ \bibinfo {author} {\bibfnamefont {S.}~\bibnamefont {Bridle}},\
  }\href {\doibase 10.1103/PhysRevD.66.103511} {\bibfield  {journal} {\bibinfo
  {journal} {Phys. Rev.}\ }\textbf {\bibinfo {volume} {D66}},\ \bibinfo {pages}
  {103511} (\bibinfo {year} {2002})},\ \Eprint
  {http://arxiv.org/abs/astro-ph/0205436} {arXiv:astro-ph/0205436 [astro-ph]}
  \BibitemShut {NoStop}%
\bibitem [{\citenamefont {{Audren}}\ \emph {et~al.}(2013)\citenamefont
  {{Audren}}, \citenamefont {{Lesgourgues}}, \citenamefont {{Benabed}},\ and\
  \citenamefont {{Prunet}}}]{MontePython}%
  \BibitemOpen
  \bibfield  {author} {\bibinfo {author} {\bibfnamefont {B.}~\bibnamefont
  {{Audren}}}, \bibinfo {author} {\bibfnamefont {J.}~\bibnamefont
  {{Lesgourgues}}}, \bibinfo {author} {\bibfnamefont {K.}~\bibnamefont
  {{Benabed}}}, \ and\ \bibinfo {author} {\bibfnamefont {S.}~\bibnamefont
  {{Prunet}}},\ }\href {\doibase 10.1088/1475-7516/2013/02/001} {\bibfield
  {journal} {\bibinfo  {journal} {JCAP}\ }\textbf {\bibinfo {volume} {2}},\
  \bibinfo {eid} {001} (\bibinfo {year} {2013})},\ \Eprint
  {http://arxiv.org/abs/1210.7183} {arXiv:1210.7183} \BibitemShut {NoStop}%
\bibitem [{\citenamefont {Lewis}\ \emph {et~al.}(2000)\citenamefont {Lewis},
  \citenamefont {Challinor},\ and\ \citenamefont {Lasenby}}]{Lewis:1999bs}%
  \BibitemOpen
  \bibfield  {author} {\bibinfo {author} {\bibfnamefont {A.}~\bibnamefont
  {Lewis}}, \bibinfo {author} {\bibfnamefont {A.}~\bibnamefont {Challinor}}, \
  and\ \bibinfo {author} {\bibfnamefont {A.}~\bibnamefont {Lasenby}},\ }\href
  {\doibase 10.1086/309179} {\bibfield  {journal} {\bibinfo  {journal}
  {Astrophys. J.}\ }\textbf {\bibinfo {volume} {538}},\ \bibinfo {pages} {473}
  (\bibinfo {year} {2000})},\ \Eprint {http://arxiv.org/abs/astro-ph/9911177}
  {arXiv:astro-ph/9911177 [astro-ph]} \BibitemShut {NoStop}%
\bibitem [{\citenamefont {{Blas}}\ \emph {et~al.}(2011)\citenamefont {{Blas}},
  \citenamefont {{Lesgourgues}},\ and\ \citenamefont {{Tram}}}]{CLASS}%
  \BibitemOpen
  \bibfield  {author} {\bibinfo {author} {\bibfnamefont {D.}~\bibnamefont
  {{Blas}}}, \bibinfo {author} {\bibfnamefont {J.}~\bibnamefont
  {{Lesgourgues}}}, \ and\ \bibinfo {author} {\bibfnamefont {T.}~\bibnamefont
  {{Tram}}},\ }\href {\doibase 10.1088/1475-7516/2011/07/034} {\bibfield
  {journal} {\bibinfo  {journal} {JCAP}\ }\textbf {\bibinfo {volume} {7}},\
  \bibinfo {eid} {034} (\bibinfo {year} {2011})},\ \Eprint
  {http://arxiv.org/abs/1104.2933} {arXiv:1104.2933} \BibitemShut {NoStop}%
\bibitem [{\citenamefont {Tzirakis}\ and\ \citenamefont
  {Kinney}(2007)}]{Tzirakis:2007bf}%
  \BibitemOpen
  \bibfield  {author} {\bibinfo {author} {\bibfnamefont {K.}~\bibnamefont
  {Tzirakis}}\ and\ \bibinfo {author} {\bibfnamefont {W.~H.}\ \bibnamefont
  {Kinney}},\ }\href {\doibase 10.1103/PhysRevD.75.123510} {\bibfield
  {journal} {\bibinfo  {journal} {Phys. Rev.}\ }\textbf {\bibinfo {volume}
  {D75}},\ \bibinfo {pages} {123510} (\bibinfo {year} {2007})},\ \Eprint
  {http://arxiv.org/abs/astro-ph/0701432} {arXiv:astro-ph/0701432 [astro-ph]}
  \BibitemShut {NoStop}%
\bibitem [{\citenamefont {Gilbert}(1995)}]{PhysRevD.52.5486}%
  \BibitemOpen
  \bibfield  {author} {\bibinfo {author} {\bibfnamefont {J.}~\bibnamefont
  {Gilbert}},\ }\href {\doibase 10.1103/PhysRevD.52.5486} {\bibfield  {journal}
  {\bibinfo  {journal} {Phys. Rev. D}\ }\textbf {\bibinfo {volume} {52}},\
  \bibinfo {pages} {5486} (\bibinfo {year} {1995})}\BibitemShut {NoStop}%
\bibitem [{\citenamefont {Anguelova}\ \emph {et~al.}(2018)\citenamefont
  {Anguelova}, \citenamefont {Suranyi},\ and\ \citenamefont
  {Wijewardhana}}]{Anguelova:2017djf}%
  \BibitemOpen
  \bibfield  {author} {\bibinfo {author} {\bibfnamefont {L.}~\bibnamefont
  {Anguelova}}, \bibinfo {author} {\bibfnamefont {P.}~\bibnamefont {Suranyi}},
  \ and\ \bibinfo {author} {\bibfnamefont {L.~C.~R.}\ \bibnamefont
  {Wijewardhana}},\ }\href {\doibase 10.1088/1475-7516/2018/02/004} {\bibfield
  {journal} {\bibinfo  {journal} {JCAP}\ }\textbf {\bibinfo {volume} {1802}},\
  \bibinfo {pages} {004} (\bibinfo {year} {2018})},\ \Eprint
  {http://arxiv.org/abs/1710.06989} {arXiv:1710.06989 [hep-th]} \BibitemShut
  {NoStop}%
\bibitem [{\citenamefont {Cai}\ \emph {et~al.}(2018)\citenamefont {Cai},
  \citenamefont {Chen}, \citenamefont {Namjoo}, \citenamefont {Sasaki},
  \citenamefont {Wang},\ and\ \citenamefont {Wang}}]{Cai:2017bxr}%
  \BibitemOpen
  \bibfield  {author} {\bibinfo {author} {\bibfnamefont {Y.-F.}\ \bibnamefont
  {Cai}}, \bibinfo {author} {\bibfnamefont {X.}~\bibnamefont {Chen}}, \bibinfo
  {author} {\bibfnamefont {M.~H.}\ \bibnamefont {Namjoo}}, \bibinfo {author}
  {\bibfnamefont {M.}~\bibnamefont {Sasaki}}, \bibinfo {author} {\bibfnamefont
  {D.-G.}\ \bibnamefont {Wang}}, \ and\ \bibinfo {author} {\bibfnamefont
  {Z.}~\bibnamefont {Wang}},\ }\href {\doibase 10.1088/1475-7516/2018/05/012}
  {\bibfield  {journal} {\bibinfo  {journal} {JCAP}\ }\textbf {\bibinfo
  {volume} {1805}},\ \bibinfo {pages} {012} (\bibinfo {year} {2018})},\ \Eprint
  {http://arxiv.org/abs/1712.09998} {arXiv:1712.09998 [astro-ph.CO]}
  \BibitemShut {NoStop}%
\bibitem [{\citenamefont {Yi}\ and\ \citenamefont {Gong}(2018)}]{Yi:2017mxs}%
  \BibitemOpen
  \bibfield  {author} {\bibinfo {author} {\bibfnamefont {Z.}~\bibnamefont
  {Yi}}\ and\ \bibinfo {author} {\bibfnamefont {Y.}~\bibnamefont {Gong}},\
  }\href {\doibase 10.1088/1475-7516/2018/03/052} {\bibfield  {journal}
  {\bibinfo  {journal} {JCAP}\ }\textbf {\bibinfo {volume} {1803}},\ \bibinfo
  {pages} {052} (\bibinfo {year} {2018})},\ \Eprint
  {http://arxiv.org/abs/1712.07478} {arXiv:1712.07478 [gr-qc]} \BibitemShut
  {NoStop}%
\bibitem [{\citenamefont {Nojiri}\ \emph {et~al.}(2017)\citenamefont {Nojiri},
  \citenamefont {Odintsov},\ and\ \citenamefont {Oikonomou}}]{Nojiri:2017qvx}%
  \BibitemOpen
  \bibfield  {author} {\bibinfo {author} {\bibfnamefont {S.}~\bibnamefont
  {Nojiri}}, \bibinfo {author} {\bibfnamefont {S.~D.}\ \bibnamefont
  {Odintsov}}, \ and\ \bibinfo {author} {\bibfnamefont {V.~K.}\ \bibnamefont
  {Oikonomou}},\ }\href {\doibase 10.1088/1361-6382/aa92a4} {\bibfield
  {journal} {\bibinfo  {journal} {Class. Quant. Grav.}\ }\textbf {\bibinfo
  {volume} {34}},\ \bibinfo {pages} {245012} (\bibinfo {year} {2017})},\
  \Eprint {http://arxiv.org/abs/1704.05945} {arXiv:1704.05945 [gr-qc]}
  \BibitemShut {NoStop}%
\bibitem [{\citenamefont {Motohashi}\ and\ \citenamefont
  {Starobinsky}(2017{\natexlab{b}})}]{Motohashi:2017vdc}%
  \BibitemOpen
  \bibfield  {author} {\bibinfo {author} {\bibfnamefont {H.}~\bibnamefont
  {Motohashi}}\ and\ \bibinfo {author} {\bibfnamefont {A.~A.}\ \bibnamefont
  {Starobinsky}},\ }\href {\doibase 10.1140/epjc/s10052-017-5109-x} {\bibfield
  {journal} {\bibinfo  {journal} {Eur. Phys. J.}\ }\textbf {\bibinfo {volume}
  {C77}},\ \bibinfo {pages} {538} (\bibinfo {year} {2017}{\natexlab{b}})},\
  \Eprint {http://arxiv.org/abs/1704.08188} {arXiv:1704.08188 [astro-ph.CO]}
  \BibitemShut {NoStop}%
\bibitem [{\citenamefont {Karam}\ \emph {et~al.}(2018)\citenamefont {Karam},
  \citenamefont {Marzola}, \citenamefont {Pappas}, \citenamefont {Racioppi},\
  and\ \citenamefont {Tamvakis}}]{Karam:2017rpw}%
  \BibitemOpen
  \bibfield  {author} {\bibinfo {author} {\bibfnamefont {A.}~\bibnamefont
  {Karam}}, \bibinfo {author} {\bibfnamefont {L.}~\bibnamefont {Marzola}},
  \bibinfo {author} {\bibfnamefont {T.}~\bibnamefont {Pappas}}, \bibinfo
  {author} {\bibfnamefont {A.}~\bibnamefont {Racioppi}}, \ and\ \bibinfo
  {author} {\bibfnamefont {K.}~\bibnamefont {Tamvakis}},\ }\href {\doibase
  10.1088/1475-7516/2018/05/011} {\bibfield  {journal} {\bibinfo  {journal}
  {JCAP}\ }\textbf {\bibinfo {volume} {1805}},\ \bibinfo {pages} {011}
  (\bibinfo {year} {2018})},\ \Eprint {http://arxiv.org/abs/1711.09861}
  {arXiv:1711.09861 [astro-ph.CO]} \BibitemShut {NoStop}%
\bibitem [{\citenamefont {Morse}\ and\ \citenamefont
  {Kinney}(2018)}]{Morse:2018kda}%
  \BibitemOpen
  \bibfield  {author} {\bibinfo {author} {\bibfnamefont {M.~J.~P.}\
  \bibnamefont {Morse}}\ and\ \bibinfo {author} {\bibfnamefont {W.~H.}\
  \bibnamefont {Kinney}},\ }\href {\doibase 10.1103/PhysRevD.97.123519}
  {\bibfield  {journal} {\bibinfo  {journal} {Phys. Rev.}\ }\textbf {\bibinfo
  {volume} {D97}},\ \bibinfo {pages} {123519} (\bibinfo {year} {2018})},\
  \Eprint {http://arxiv.org/abs/1804.01927} {arXiv:1804.01927 [astro-ph.CO]}
  \BibitemShut {NoStop}%
\end{thebibliography}%

\end{document}